\documentclass{aa}
\usepackage[utf8]{inputenc}
\usepackage[varg]{txfonts}
\usepackage{multirow}
\usepackage{lmodern}
\usepackage{natbib}
\usepackage[english]{babel}
\usepackage{physics}
\usepackage{wasysym}
\usepackage{multicol}
\usepackage{amsmath}
\usepackage[section]{placeins}
\usepackage{booktabs}
\usepackage{caption}
\newcommand{\citeaposauthor}[1]{\citeauthor{#1}'s}
\newcommand{\citeapos}[1]{\citeauthor{#1}'s (\citeyear{#1})}

\newcommand{\nint}[1]{\ensuremath{\lfloor#1\rceil}}

\usepackage{hyperref}

\hypersetup{
  colorlinks   = true,
  urlcolor     = blue,
  linkcolor    = blue,
  citecolor    = blue
}
\usepackage{etoolbox}

\makeatletter

\pretocmd{\NAT@citex}{%
  \let\NAT@hyper@\NAT@hyper@citex
  \def\NAT@postnote{#2}%
  \setcounter{NAT@total@cites}{0}%
  \setcounter{NAT@count@cites}{0}%
  \forcsvlist{\stepcounter{NAT@total@cites}\@gobble}{#3}}{}{}
\newcounter{NAT@total@cites}
\newcounter{NAT@count@cites}
\def\NAT@postnote{}

\def\NAT@hyper@citex#1{%
  \stepcounter{NAT@count@cites}%
  \hyper@natlinkstart{\@citeb\@extra@b@citeb}#1%
  \ifnumequal{\value{NAT@count@cites}}{\value{NAT@total@cites}}
    {\ifNAT@swa\else\if*\NAT@postnote*\else%
     \NAT@cmt\NAT@postnote\global\def\NAT@postnote{}\fi\fi}{}%
  \ifNAT@swa\else\if\relax\NAT@date\relax
  \else\NAT@@close\global\let\NAT@nm\@empty\fi\fi
  \hyper@natlinkend}
\renewcommand\hyper@natlinkbreak[2]{#1}

\patchcmd{\NAT@citex}
  {\ifNAT@swa\else\if*#2*\else\NAT@cmt#2\fi
   \if\relax\NAT@date\relax\else\NAT@@close\fi\fi}{}{}{}
\patchcmd{\NAT@citex}
  {\if\relax\NAT@date\relax\NAT@def@citea\else\NAT@def@citea@close\fi}
  {\if\relax\NAT@date\relax\NAT@def@citea\else\NAT@def@citea@space\fi}{}{}

\makeatother

\begin{document}

\title{The distribution of lenticular galaxies in the phase space of present-day galaxy cluster regions}

  \author{M.\ Gort
          \inst{1}\fnmsep\thanks{Email: mathias.gort173@gmail.com}, 
          J.\ L.\ Tous\inst{2}
          \and
          J.\ M.\ Solanes\inst{1,3}
          }

   \institute{Institut de Ci\`encies del Cosmos (ICCUB), Universitat de Barcelona. C.\ Mart\'{\i}  i Franqu\`es, 1, E-08028 Barcelona, Spain
       \and
         School of Physics and Astronomy, University of Southampton, Highfield, Southampton SO17 1BJ, UK
        \and
             Departament de F\'\i sica Qu\`antica i Astrof\'\i sica, Universitat de Barcelona. C.\ Mart\'{\i}  i Franqu\`es, 1, E-08028 Barcelona, Spain
             }

   \date{v.1.10 04/04/2025. Accepted.}

\abstract
{Lenticular (S0) galaxies are ubiquitous in both high- and low-density environments where diverse evolutionary mechanisms operate. Consequently, studying their distribution and properties across both the dense, virialized cluster cores and their sparser surrounding secondary infall regions can provide key insights into the still-debated processes driving their evolution. In this work, we investigated the environmental impact of cluster regions on the evolution of present-day S0 galaxies, focusing on their distinct quiescent and star-forming (SF) subpopulations. We selected a sample of nearby cluster regions by crossmatching optical and X-ray data and extract a subset of 14 systems with maximally relaxed cores by applying strict virialization and substructure tests. A projected phase space (PPS) diagram was then generated from the stack of maximally relaxed clusters up to 3 virial radii to assess the locations of quiescent and SF S0s and their cluster infall histories. Additionally, we compared the radial line-of-sight velocity dispersion (VDLOS) and specific star-formation rate (SSFR) profiles for the different S0 subpopulations, using other Hubble types as benchmarks. Our study shows that quiescent S0s, the dominant class in the entire cluster region, concentrate preferentially at low radii in the PPS diagram, while their SF counterparts are more abundant in the outskirts. Despite this segregation, quiescent and SF S0s exhibit similar VDLOS profiles in the dynamically relaxed cluster core ---indicating an advanced stage of dynamical relaxation---, but that resemble those of late-type galaxies beyond the virial radius. This finding, combined with the distinct PPS distributions of both S0 subpopulations, which lead to mean infall times $\sim 1$ ~Gyr longer for quiescent S0s but that are shorter than those expected for ancient infallers, suggests that a substantial fraction of S0s present in the core region arrive via secondary infall. We also find evidence in the radial SSFR profiles that star formation in S0s begins to decline well beyond the virialized core, likely due to preprocessing in infalling groups. Overall, our results support a delayed-then-rapid quenching scenario for SF S0s in cluster regions, where their centrally concentrated star formation persists for an extended period before abruptly ending ($\lesssim 0.1$ Gyr) after their first pericenter passage.}
    

\titlerunning{Distribution of S0 galaxies in the phase space of galaxy clusters}
\authorrunning{M.\ Gort et al.}
\maketitle

\section{Introduction}

Lenticular galaxies (S0s) occupy a distinct position in the Hubble classification scheme, bridging the gap between elliptical and spiral galaxies. While in terms of the stellar population they are closer to elliptical galaxies, their structural properties align with disk galaxies, featuring thicker, lens-shaped disks devoid of spiral arms. Additionally, unlike other morphological types, S0 galaxies are abundant in both high- and low-density galactic environments, suggesting a complex ---and potentially diverse--- set of formation and evolutionary pathways \citep{Wilman&Erwin2012}. The distribution of S0 galaxies in the Local Universe adheres to the morphology-density relation first outlined by \citet{Dressler1980}. This relation highlights that, despite their ubiquity, S0s preferentially inhabit the densest cosmic environments, galaxy groups and, particularly, the dynamically relaxed cores of galaxy cluster regions. Consequently, examination of the differences in the properties of S0s between the virialized and non-virialized regions of present-day galaxy clusters, can offer valuable insights into the still poorly understood physical processes that shape this intriguing morphological class.

In recent years, numerical simulations of galaxy interactions have concluded that S0s can form through minor mergers between disk galaxies and small satellites, as well as major mergers between gas-rich, star-forming (SF) spiral pairs \citep{Querejeta2015, Tapia2017, Eliche-Moral2018}. Additionally, it is widely accepted that lenticular galaxies can also form through hydrodynamical interactions between the galactic interstellar medium (ISM) and the surrounding intragroup or intracluster medium (IGM and ICM, respectively) via mechanisms such as ram pressure stripping (RPS) capable of depleting the neutral gas of the disks of spiral galaxies when they move at large speeds in these high-density environments \citep{Gunn&Gott1972, Nulsen1982}. Other transport processes \citetext{thermal evaporation and turbulent stripping \citealt{Nulsen1982}}, tidal interactions, galaxy harassment \citep{Moore1999}, and/or starvation \citep{Larson1980, Tinsley1980}, can also contribute to the morphological transformation of spiral galaxies into S0s in groups and clusters of galaxies \citep{Solanes&Salvador-Sole1992, Dressler1997, Fasano2000, Quilis2000, Crowl2008, Poggianti2009, D'Onofrio2015}. 

Observational studies, both photometric and spectroscopic, have reinforced the notion that S0 galaxies do not adhere to a single evolutionary pathway. These studies have identified at least two subpopulations within this morphological class, each with significantly distinct intrinsic properties \citep{Xiao2016, Fraser-McKelvie2018, Deeley2020, DominguezSanchez2020, Coccato2022, Rathore2022}. Supporting this view, \citet{Tous2020}, have conducted a comprehensive analysis of the variance of the optical integral spectra of 68,043 S0 galaxies with $z\leq 0.1$ retrieved from the Main Galaxy Sample of the Sloan Legacy Survey \citep{Strauss2002}, uncovering a broad range of star formation activity within this Hubble type. By applying Principal Component Analysis (PCA) to these spectra, the authors have demonstrated that their projections into the latent space defined by the first two principal components (PC1 and PC2) basically cluster into two well-separated regions encompassing objects with statistically inconsistent physical properties. Based on these findings, \citeauthor{Tous2020} have categorized S0 galaxies into three different spectral classes according to their positions in this subspace. The most populous of these, called the "passive sequence" (S0--PS), consists of systems with quiescent optical spectra and strongly correlated PCs. This group forms a narrow yet very densely populated diagonal band that accounts for approximately $70\%$ of all lenticular galaxies in the local Universe. The second class, referred to as the "active cloud" (S0--AC), consists of S0 galaxies with emission lines. These objects occupy a less crowded and much more extended region adjacent to the PS, representing about $25\%$ of the S0s\footnote{As discussed in Appendix A of \citet{Tous2020}, a key advantage of this PCA-based classification of the single-fiber optical spectra of S0s is its robustness against aperture and internal extinction effects.}. Separating these two major classes is a small, narrow "transition region" (S0--TR), that encompasses the remaining $\sim 5\%$ of the S0 galaxies with intermediate spectral properties. Overall, the members of the S0--AC population compared to their PS counterparts are less massive, more luminous with less concentrated light profiles, younger, bluer, metal-poorer, and preferentially reside in lower density environments. In terms of their optical spectra, S0--PS occupy essentially the same region in the PC1--PC2 subspace as elliptical galaxies, while the area covered by the S0--AC substantially overlaps with the region defined by spiral galaxies in this same diagram. In fact, as we will also show in the present analysis, members of the S0--AC class often exhibit levels of star formation similar to those of intermediate Hubble types \citep{Kaviraj2007,Tous2020}, suggesting that, contrary to common assumptions, the local population of S0s is not entirely composed of "red and dead objects". These findings about the star formation activity of lenticular galaxies and their diverse formation pathways are supported by recent state-of-the-art hydrodynamic cosmological simulations \citep[e.g.,][]{Deeley2021}.

This work aims to advance our understanding of the evolutionary pathways of S0 galaxies by examining their spatial and kinematic distributions within the galaxy cluster regions that populate the Local Universe ($z\lesssim 0.1$). These cluster fields are defined to encompass both the central, dynamically relaxed core, where a deep gravitational potential fosters galaxy harassment and strong hydrodynamic interactions, and the surrounding secondary infall region, where galaxy groups dominate. In this outer region, prolonged gravitational interactions facilitate tidal encounters and mergers, which, along with mild forms of hydrodynamic interactions, are expected to play a crucial role in preprocessing galaxies before they enter the cluster core and become fully virialized. Building on the work of \citet{Rhee2017}, we investigate in this study which classes of S0 galaxies populate the distinct regions of the projected phase space (PPS) of galaxy clusters\footnote{The observational 3D projection of the full 6D phase-space distribution function $f(\mathbf{r},\mathbf{v})$, which encodes the complete kinematic and spatial structure of these galaxy systems.}, defined by the projected clustercentric distances and line-of-sight (LOS) systemic velocities of member galaxies. By leveraging the predictive power of this diagram and examining trends in the line-of-sight velocity dispersion (VDLOS) and star formation activity as a function of clustercentric distance ---using for the latter property mass-independent metrics such as the specific star-formation rate (SSFR) and the distance to the Main Sequence of star formation, $\Delta\mbox{MS}$---, we aim to gain deeper insights into the evolutionary pathways and environmental mechanisms shaping the S0 population.

The structure of this paper is as follows. Sect.~\ref{sec:data} introduces the dataset used in our analysis. In Sect.~\ref{sec:sampleselection}, we describe the criteria and methodology employed to identify cluster regions with the most dynamically relaxed cores. We then construct in Sect.~\ref{sec:results} a stacked sample of these systems, which is used in Sect.~\ref{sec:PPSS0MRsample} to investigate the positioning of different S0 classes within the projected phase space (PPS) of dense environments. This stack of clusters with virialized cores is also used in Sect.~\ref{sec:profiles}, to derive radial VDLOS and SSFR profiles for the various S0 subpopulations, and compare them with those of other galaxy types. The broader implications of our results are discussed in Sect.~\ref{sec:discussion} and a summary of our main conclusions is presented in Sect.~\ref{sec:conclusions}. This manuscript also includes Appendix~\ref{app:plotsclusters}, where we present PPS and sky projections plots of the studied cluster regions, and Appendix~\ref{extra_material}, which provides supplementary figures that visually support the discussions in the main text. For all calculations that depend on the cosmological model, a standard flat Lambda Cold Dark Matter ($\Lambda$CDM) cosmology with $H_0 = 70\;\mbox{km}\;\mbox{s}^{-1}\;\mbox{Mpc}^{-1}$, $\Omega_{\mathrm{m,0}}=0.3$, and $\Omega_{\Lambda,0}=0.7$ is adopted.

\section{Data}
\label{sec:data}

Here, we outline the main characteristics of the data catalogs used to select the present-day galaxy cluster fields and S0 galaxies analyzed in this study.

\subsection{Galaxy clusters from optical data}
\label{sec:clustersandgalaxies}

Our local galaxy cluster regions are drawn from the group catalog of \citet{Tempel2017}, which defines groups independent of mass or membership size, with as few as two galaxies being the minimum number necessary to be considered a group. These authors provide three catalogs, all based on the 12th data release of the Sloan Digital Sky Survey \citetext{\citealt{Eisenstein2011,Alam2015}; hereafter SDSS-DR12}. The first of them consists of all galaxies that were considered in the survey, regardless if they are members of groups or isolated. Only galaxies included in the main contiguous area of the SDSS survey with spectroscopic and photometric data and apparent Petrosian $r$-band magnitude brighter than $17.77$ (i.e.,\ the Main SDSS Legacy Survey), as well as with $z\le 0.2$ are included in this catalog. The total number of objects is $584,449$. The second catalog lists all groups they identify (see below), while the third dataset collects all galaxy systems that show evidence of merging.

To find groups, \citeauthor{Tempel2017} adopted the Friends-of-Friends (FoF) algorithm \citep{Turner1976, Beers1982, Zeldovich1982, Barnes1985} with iterative membership refinement procedures that assumed a Navarro-Frenk-White (NFW) total matter density profile \citep{Navarro1995} for these galaxy systems. With this approach, they identified $88,662$ galaxy groups with two or more members, including 498 potential merging systems, across the contiguous area of the Legacy Survey. Their data indicate that $49\%$ of galaxies belong to one of the detected groups. Additionally, while flux limitations result in a scarcity of the richest groups at large distances, \citeauthor{Tempel2017}'s data also show that the maximum number of group members remains roughly constant up to $z\simeq 0.1$. This ensures a uniform sampling of galaxy cluster regions in the Local Volume, making this dataset perfectly suited for the purposes of the present study.

\subsection{X-ray galaxy clusters}
\label{sec:xrayclusters}

The presence of strong, extended X-ray emission in a galaxy cluster is a clear indication of a hot ICM, signifying that their central regions are relaxed and virialized. In the most evolved clusters, the X-ray emission peak is generally not associated with any individual galaxy, except possibly with the BCG, and typically exhibits little to no offset from the system's center of mass. To ensure that the clusters selected for this study are dynamically evolved structures, two catalogs of X-ray selected galaxy clusters are combined with the optical cluster data described in the previous section.

These two catalogs are from the X-Class survey \citep{Koulouridis2021} and the ROSAT All-Sky Survey (RXGCC) \citep{Xu2021}. The X-Class cluster catalog comprises a total of $1,646$ systems with redshifts extending up to $z\sim 1.5$, though the majority are concentrated around $z\sim 0.1$. Among these, 297 clusters ($\sim 19\%$ of the total sample) satisfy the criterion of being located in the Local Universe. In contrast, the RXGCC catalog contains a total of 944 X-ray clusters, of which 619 ($66\%$) lie within our redshift limit of $z \leq 0.1$. The combined total of X-ray clusters retrieved from both catalogs is therefore 916. However, as these surveys cover overlapping regions of the sky, some galaxy systems may be duplicated. To address this, we cross-matched the two cluster subsets by comparing their X-ray center coordinates and redshifts, allowing for maximum uncertainties of 3 arcmin in position (a small fraction of the typical spatial extent of these clusters in the sky) and 0.005 in redshift (corresponding to the mean redshift uncertainty in the RXGCC catalog). This process identified 120 duplicate clusters, which were then removed from the X-Class catalog and retained only in the RXGCC dataset to enhance uniformity. Thus, our final X-ray cluster sample, constructed by merging the two catalogs, consists of 796 unique clusters.

\subsection{Lenticular galaxies}
\label{sec:lenticulars}

The source of the present-day S0 galaxy sample is the catalog compiled by \citet{Tous2020}, which includes a total of $68,043$ galaxies of this Hubble type with $z \leq 0.1$. This dataset provides key information such as the galaxies' location, redshift, and spectral class (PS, TR or AC), as well as a comprehensive list of their physical properties sourced from the literature. To identify galaxies of lenticular morphology, \citet{Tous2020} utilized the catalog from \citet{DominguezSanchez2020}, who applied machine learning techniques to the Main SDSS Legacy Survey to assign $T$-type values to galaxies. They also calculate the probability of being S0 ($P_{\rm S0}$) for objects with $T \leq 0$, to help distinguish between E and S0 galaxies. \citeauthor{Tous2020} classified any galaxy with \nint{T}$\,\leq 0$ and $P_{\rm S0} > 0.7$ as an S0, selecting this conservative probability threshold to ensure a high purity for the S0 sample.

\section{Selection of cluster regions with dynamically evolved cores along with their galaxy members}
\label{sec:sampleselection}

To investigate the distribution of S0 galaxies in the PPS of cluster regions, it is fundamental to select galaxy associations with fully evolved cores free of substructure, ensuring that the positions of the galaxies in the PPS can be interpreted correctly. In this section, we outline the steps taken to obtain a final sample of galaxy clusters that exhibit no obvious dynamical disturbances in their central region, along with their galaxy members.

\subsection{Preliminary constraints}
\label{sec:selectionfinalclusters}

To begin, we applied the following filters to the optical galaxy associations identified by \citet{Tempel2017} (see Sect.~\ref{sec:clustersandgalaxies}):

\begin{itemize}
    \item The system's redshift cannot exceed $0.1$, ensuring that it is located in the Local Universe.
    \item The system's virial mass must be larger than $10^{14}~\mbox{M}_{\odot}$, selecting only bona fide galaxy clusters.
    \item The system's radius, defined as the distance from the system's center to the farthest member galaxy in the plane of the sky, must be greater than $R_{200}$, as measured by \citeauthor{Tempel2017}, but not exceed $5$ Mpc, eliminating unrealistically small or large FoF galaxy associations.
    \item The minimum number of galaxy members within the innermost $1$ Mpc radius of the system has to be at least 20, ensuring sufficient statistical representation.
    \item The system must not be listed in the third catalog of \citet{Tempel2017}, which identifies potentially merging galaxy associations. 
\end{itemize}

An additional refinement of this cluster sample was performed by crossmatching the optical systems that pass the initial filtering with our X-ray selected clusters (see Sect.~\ref{sec:xrayclusters}). For this task, we employed the same strategy used for crossmatching the RXGCC and X-Class X-ray cluster catalogs, comparing the sky coordinates and redshifts of the cluster centers allowing for absolute errors of 3 arcmin in the coordinates and $0.005$ in the redshift. 

This final criterion results in a preliminary sample of 37 galaxy cluster regions that exhibit tentative signs of harboring dynamically evolved cores. However, these cluster cores must undergo additional testing (see Sects.~\ref{sec:testvir} and \ref{sec:testsubstructure}) to confirm their relaxed dynamical status, after which the selection will be refined to a final subset comprising the most suitable cluster fields for this study.

At this stage, it is important to remark that \citet{Tempel2017} used $R_{200}$ and the associated total mass, $M_{200}$\footnote{$R_{200}$ is the clustercentric radius and $M_{200}$ is the total mass enclosed within that radius, where the mean density of the system exceeds the cosmic average by a factor of 200.}, as proxies for the corresponding virial values. Although this is a standard procedure in the literature, both parameters underestimate the true virial values, since the density contrast of galaxy clusters in a standard flat $\Lambda$CDM cosmology is actually very close to 100\footnote{More precisely, 100.14 for the adopted cosmology.} \citep[see, e.g.,][]{Mamon2004,DS2010}. For this reason, in the next section, we derive more accurate estimates of the virial magnitudes. We also note that while \citeauthor{Tempel2017} assumed this same cosmology, they choosed $H_0 = 67.8\,\text{km~s}^{-1}\,\text{Mpc}^{-1}$. In this case, however, such a slight discrepancy with our adopted $H_0$ value has been ignored throughout the present study.

\subsection{Refinement of cluster properties and membership}
\label{sec:selectionfinalmembers}

Investigating the PPS of cluster regions requires a thorough identification of their galaxy members. Notably, clusters predominantly grow through the accretion of galaxy groups rather than individual galaxies \citep{Berrier2009, McGee2009}. As a result, all galaxies in the outskirts of a cluster region should be regarded as part of the same cluster field, regardless of whether they are associated with smaller groups undergoing accretion. Since each galaxy in the \citeapos{Tempel2017} catalogs is assigned to only one group, some galaxies in the secondary infall region of a cluster may not be classified as part of that cluster field, despite being physically associated with it. To ensure a comprehensive membership definition for our sample of cluster regions, this potential misclassification must be addressed, as the validity of the results depends to a great extent on including all galaxies inhabiting cluster outskirts. This led us to disregard the membership assignment provided in these catalogs and reformulate this classification using our own criteria.

To achieve this goal, we first applied a recursive method to recalculate the central coordinates, redshifts, and velocity dispersions of our preselected cluster fields. To begin with, we estimated the redshift of each cluster region, $z_{\mathrm{clu}}$, by averaging the redshifts of all galaxies identified as cluster members in \citet{Tempel2017} within a projected radius of $1$~Mpc from the original center. Next, we applied a $\pm 4\sigma$-clipping technique to remove any possible contamination of foreground or background galaxies. The galaxies that kept their status of cluster members were then used to re-estimate the central coordinates and redshift of the cluster, using the robust biweight estimator for location from \cite{Beers1990}:
\begin{equation}
\label{eq:estimofloc}
    C_{BI} = M + \frac{\sum_{|u_i|<1}(x_i - M)(1 - u_i^2)^2}{\sum_{|u_i|<1}(1 - u_i^2)^2}\;,
\end{equation}
where $x_i$ is the $i$th smallest datum (sky coordinates, $\mbox{R.A.}$ and $\mbox{Dec.}$, or redshift, $z$), $M$ is the median of the values of the corresponding sample, and the $u_i$ are inferred from
\begin{equation}
    u_i = \frac{x_i - M}{k \cdot \textit{MAD}}\;.
\end{equation}

\noindent In the last equation, the best efficiency for location estimation is provided by a tuning constant $k=6$, which uses data up to four standard deviations from the central location \citep{Mosteller1977}, while $MAD\equiv\mbox{median}(\left| x_i - M \right|)$ is the median absolute deviation from the sample median. \cite{Beers1990} also provide a robust estimator of scale 
\begin{equation}
\label{eq:estimofscale}
    S_{BI} = \sqrt{N} \frac{\left[ \sum_{|u_i|<1}(x_i - M)^2 (1 - u_i^2)^4 \right]^{1/2}}{\left| \sum_{|u_i|<1}(1 - u_i^2) (1 - 5 u_i^2) \right|}\;,
\end{equation}
which was used to determine the cluster velocity dispersion, $\sigma_{\mathrm{clu}}$, from the LOS velocities in the cluster rest frame of the same $N$ galaxies used to calculate the new center coordinates and redshift of the cluster, which are given by
\begin{equation}
\label{eq:vlos}
    v_{\mathrm{LOS},i}  = c \frac{z_i - z_{\mathrm{clu}} }{1 + z_{\mathrm{clu}}}\;, \ \ \ \ \ \ \ \ \ \ i=1,\ldots,N.
\end{equation}

All these estimators were further refined by recalculating their values until convergence, which is achieved after a few (typically three) iterations. It is important to remark that, in typical virialized galaxy clusters, the member galaxies within the innermost 1 Mpc provide a robust sample of the overall cluster dynamics while minimizing contamination from interlopers. For this reason, the values calculated from the above recursive procedure should be representative of the entire cluster region, particularly its dynamically evolved core.

The next step was to estimate the cluster virial radius, $R_{\mathrm{vir}}$, from the following equation:

\begin{equation}
\label{eq:rvir}
\begin{split}
      \left[\frac{R_{\mathrm{vir}}}{\mbox{Mpc}}\right] = 1.43\cdot 10^{-3} \left[\frac{\sqrt{3}\sigma_{\mathrm{clu}}}{\mbox{km}\;\mbox{s}^{-1}}\right] & \left[\frac{\Omega_{\rm{m,0}} \Delta_{\mathrm{vir,0}}}{200}\right]^{-1/2}\cdot \\ & \ \ \ \ \ \ (1+z_{\mathrm{clu}})^{-3/2}\;,
\end{split}
\end{equation}

\noindent with $\Delta_{\mathrm{vir,0}} \simeq 337$ the virial overdensity at $z \sim 0$ \citep{Bryan&Norman1998} for the adopted cosmology. The cluster virial mass, $M_{\mathrm{vir}}$, is then simply given by
\begin{equation}
    \left[\frac{M_{\mathrm{vir}}}{\text{M}_{\odot}}\right] = \frac{4 \pi}{3}R_{\mathrm{vir}}^3 100\rho_{\mathrm{crit,0}}\;,
\end{equation}

\noindent with the current value of the critical density $\rho_{\mathrm{crit,0}} = 1.39\cdot 10^{11}~\text{M}_{\odot}~\text{Mpc}^{-3}$.

\subsection{Further refinement of membership using caustics}
\label{sec:caustics}

To further refine cluster membership, we computed the caustic lines in the PPS of each cluster. This method imposes physical boundaries based on the cluster’s escape velocity, which is derived from the total enclosed mass at each radius. In projection, these boundaries form a characteristic trumpet-shaped structure. Only galaxies located within the caustic lines ---meaning those with LOS velocities lower than the escape velocity at their projected clustercentric distance--- are considered bona fide cluster members. A key advantage of this approach is that caustic boundaries can be traced out to several virial radii, as their determination depends solely on the total interior mass rather than the dynamical state of the system. The only underlying assumption is spherical symmetry, which may not hold for individual clusters, particularly in their outskirts, where deviations from sphericity are relatively common (but see Sect.~\ref{sec:PPSS0MRsample}).

The caustics were inferred by following the technique of \cite{Diaferio1997}, where the escape velocity at each projected radius, $r_\bot$, was calculated from the equation
\begin{equation}
    v_{\mathrm{esc}}(r_\bot) = \pm \left\{ -2 \Phi(r) \frac{1-\beta(r)}{3-2\beta(r)} \right\}^{\frac{1}{2}}_{r=r_\bot}\;.
    \label{vesc}
\end{equation}

\noindent In Eq.~(\ref{vesc}), $\Phi(r)$ is the gravitational potential, which is corrected by a function that depends on the velocity anisotropy profile $\beta(r)$. Cosmological simulations show, that galaxy motions in the centers of relaxed clusters are mainly random, implying isotropy and therefore $\beta(0) \sim 0$, while near the virial radius, motions become anisotropic, causing $\beta$ to increase. A good fit to the average anisotropy profile of cluster-mass cosmological halos is provided by the model of \citet{Tiret2007}:
\begin{equation}
    \beta(r) = \beta_{\infty} \frac{r}{r + r_{-2}},
\end{equation}
with $\beta_{\infty} = 0.5$ and $r_{-2}$ the radius of logarithmic slope $-2$ in the mass density profile. For a NFW mass density profile $r_{-2}$ coincides with the profile scale length $r_s$ (see below), which is related to the virial radius by the concentration parameter $C\equiv R_{\mathrm{vir}}/r_s$. For the calculation of the caustics, we assumed $C=6$ for all clusters, which is the value characteristic of cold dark matter halos with total masses on the order of a few $\times 10^{14-15}~\mbox{M}_{\odot}$ \citetext{\citealt{Maccio2008}; see Sect.~\ref{sec:resfinalclusters}}.

The last component required to derive the caustic lines is the gravitational potential $\Phi(r)$. Assuming that the total matter density profiles of our galaxy cluster regions are well approximated by a NFW profile,
\begin{equation}
\label{eq:rho}
    \rho(r) = \frac{\rho_s}{r/r_s (1 + r/r_s)^2}\;,
\end{equation}
where
\begin{equation}
    \rho_s = \frac{C^3}{3} \left[ \ln(1+C) - \frac{C}{1+C} \right]^{-1} \Bar{\rho}_{\mathrm{clu}}\;,
\end{equation}
with $\bar{\rho}_{\mathrm{clu}}=100 \rho_{\mathrm{crit,0}}$, the integral of the mass density profile in Eq.~(\ref{eq:rho}) yields the total halo mass, which with the aid of Poisson's equation can be used to derive the radial dependence of the gravitational potential:
\begin{equation}
    \Phi(r) = \frac{4 \pi G \rho_s r_s^3}{r} \ln \left(1 + \frac{r}{r_s} \right).
\end{equation}

In this study, we focused on galaxies located within the caustic lines, extending out to $3\,R_{\mathrm{vir}}$ from the cluster center. This selection ensures a fair representation of the extended cluster environment beyond the virialization sphere, where galaxies are still infalling toward the cluster core with a nonzero mean radial velocity\footnote{A recent study based on a large set of IllustrisTNG clusters with $M_{200}>10^{14}M_\odot$ and $0.01\leq z\lesssim 1.0$ by \citet{Pizzardo2024} has determined that the turnaround radius of galaxy clusters, defined as the clustercentric distance where galaxies decouple from the Hubble flow, has a typical $z$-independent value of $4.8\,R_{200}\sim 4\,R_{\mathrm{vir}}$.}.

\subsection{Identification of the brightest cluster galaxies}
\label{sec:identifyBCGs}

In fully virialized clusters, the BCGs\footnote{True BCGs are cluster galaxies which not only are ranked highest in terms of luminosity, but also are distinguished by a significant magnitude gap relative to the second-brightest member.} are expected to occupy a prominent position near the peak of X-ray emission, which in turn frequently coincides with the center of mass of the cluster. When this alignment is absent, it may indicate incomplete relaxation, often also evidenced by the presence of substructure in the core region.

To ascertain the presence of a truly dominant galaxy within the systems that conform our sample of preselected cluster regions, we followed a procedure consistent with that outlined in \citet{Bilata-Woldeyes2025}. Specifically, we calculated the differences in absolute $r$-band magnitudes between the first and second-brightest galaxies, $\Delta{\cal{M}}_{2-1}$, and between the second and third ones, $\Delta{\cal{M}}_{3-2}$, located within $1\,R_{\mathrm{vir}}$. Using these magnitude gaps, we categorized our clusters into single-BCG systems, when a distinct dominant galaxy with $\Delta{\cal{M}}_{2-1}\geq 0.45$ mag is identified (i.e.,\ when the BCG is at least $50\%$ brighter than the second ranked galaxy). On the other hand, double-BCG clusters are those for which $\Delta{\cal{M}}_{2-1} < 0.45$ mag but $\Delta{\cal{M}}_{3-2} \ge 0.45$ mag, indicating the presence of a pair of bright galaxies that stand out from the rest of the cluster members. The remaining clusters for which both $\Delta{\cal{M}}_{2-1}$ and $\Delta{\cal{M}}_{3-2}$ are less than 0.45 mag were considered non-BCG clusters, as they lack a truly dominant object. Although these latter systems are expected to exhibit lower levels of dynamical relaxation compared to single- or double-BCG clusters, we did not exclude them from our analysis as long as they pass the tests for virialization and substructure described in the next two sections. These tests were applied to the full sample of candidate cluster fields to select those with the most dynamically relaxed cores for our study.

\subsection{Test of virialization}
\label{sec:testvir}

Unrelaxed systems often exhibit discrepancies among the various indicators used to determine their center of mass, including the overall distribution of galaxy members, the peak of X-ray emission, and the sky coordinates and recession velocity of the BCG. The indicator test \citep{Gebhardt1991} focuses on verifying if the velocity of the BCG matches the velocity distribution of the cluster member galaxies. The brightest of the galaxies is used instead of the center of X-ray emission because the ICM gas is collisional and therefore relaxes faster than the galaxies. This test in particular estimates the probability of drawing the observed BCG velocity from the distribution of LOS velocities of cluster members, which is given by Eq.~(\ref{eq:vlos}) but now with $N=N_{\mathrm{clu}}$, the total number of cluster members within $1\,R_{\mathrm{vir}}$.

The "$Z$-score" for the BCG is then inferred from the expression
\begin{equation}
    Z = \frac{v_{\mathrm{BCG}} - v_{\mathrm{clu}}}{\sigma_{\mathrm{clu}}}\;,
\end{equation}
with $v_{\mathrm{BCG}}$ the line-of-sight velocity of the BCG in the cluster rest frame and $v_{\mathrm{clu}}=c\!\cdot\!z_{\mathrm{clu}}$. 

To test if a BCGs velocity is consistent with the velocity distribution of its host cluster, confidence intervals of the $Z$-score are calculated by using bootstrap resampling. In this work, we conducted $10,000$ resamplings for each BCG and considered an offset of the BCG velocity significant when the $90\%$ confidence interval\footnote{A $90\%$ confidence interval is chosen for consistency with the test of substructure in Sect.~\ref{sec:testsubstructure}.} does not bracket zero. 

\begin{table}[]
    \centering
    \caption{Results of virialization and substructure tests for the MR clusters.}
    \label{tab:resvirandsub}
    \begin{tabular}{ccc}
    \toprule
        ID$^a$ & $Z$-score & $P(\Delta)$ \\ 
    \midrule
        205 & $-$0.04 ($-$0.24, 0.28) & 0.71 \\
        262 & $-$0.23 ($-$0.36, 0.30) & 0.18 \\
        292 & $-$0.25 ($-$0.32, 0.32) & 0.39 \\
        571 & $-$0.15 ($-$0.31, 0.37) & 0.11 \\
        734 & $+$0.10 ($-$0.38, 0.38) & 0.97 \\
        804 & $-$0.31 ($-$0.34, 0.48) & 0.96 \\
        2236A$^b$ & $-$0.27 ($-$0.43, 0.41) & 0.92 \\
        2236B$^b$ & $-$0.17 ($-$0.43, 0.41) & 0.92 \\
        2316 & $-$0.04 ($-$0.57, 0.36) & 0.74 \\
        2496 & $-$0.01 ($-$0.28, 0.26) & 0.84 \\
        2620 & $+$0.02 ($-$0.44, 0.25) & 0.32 \\
        5183 & $-$0.06 ($-$0.33, 0.34) & 0.41 \\
        5567 & $-$0.26 ($-$0.29, 0.30) & 0.83 \\
        5905 & $+$0.21 ($-$0.28, 0.36) & 0.20 \\
        8062 & $+$0.07 ($-$0.24, 0.17) & 0.26 \\ 
        \midrule
        Mean$^c$ & $-$0.09 ($-$0.34, 0.33) & 0.57 \\
        \bottomrule
        \noalign{\smallskip}
    \end{tabular}
    \vspace{3pt}
    \parbox{\linewidth}{\footnotesize
    $^a$ Cluster ID in \citeaposauthor{Tempel2017} (\citeyear{Tempel2017}) catalog. \\
    $^b$ Cluster 2236 appears listed twice (A,B) with different $Z$-scores, as it hosts two nearly equally luminous dominant galaxies that pass the indicator test (see Sect.~\ref{BCGresults}).\\
    $^c$ Mean of the values in the columns.}
\end{table}

\subsection{Test of substructure}
\label{sec:testsubstructure}

A common tool used in the literature for the detection of substructure in clusters and, therefore, with the potential of being predictive in terms of the dynamical state of a cluster is the Dressler-Shectman (DS) test \citep{Dressler1988}. Having substructure means that clusters can be split into multiple subunits according to the positions and velocities of their member galaxies. Only galaxies within $1\,R_{\mathrm{vir}}$ were considered for this test, as only the cluster cores are expected to be in approximate dynamical equilibrium.

To estimate the substructure within a cluster, the DS test takes into account the spatial and velocity information of each member galaxy. With this data the $\Delta$ statistic is computed
\begin{equation}
\label{eq:dstest}
    \Delta = \sum_{i=1}^{N_{\mathrm{clu}}} \delta_i = \sum_{i=1}^{N_{\mathrm{clu}}} \sqrt{\frac{11}{\sigma_{\mathrm{clu}} ^2} \left[ (v_{\mathrm{LOS},i}  - v_{\mathrm{clu}} )^2 + (\sigma_{\mathrm{nn},i} - \sigma_{\mathrm{clu}} )^2 \right]}\;,
\end{equation}
\noindent where the $\delta_i$ are local estimates of the  degree of substructure for each galaxy, with $v_{{\mathrm{LOS},i}}$ the LOS velocity of each galaxy (eq.~[\ref{eq:vlos}]), and $\sigma_{{\mathrm{nn},i}}$ the local velocity dispersion inferred from its 10 nearest neighbors, resulting in the factor 11 in front of the equation, which is the standard choice for this test. 

To determine the actual degree of substructure, the computed $\Delta$ value is compared with the distribution of values of the same parameter obtained in a number of Monte Carlo randomizations \citep{Halliday2004, Rumbaugh2013} of the cluster data, by shuffling the galaxy velocities while maintaining their spatial distribution. A large value of $\delta_i$ for a given galaxy implies a high probability for it to be located in a spatially compact subsystem. $P(\Delta)$ then measures the probability that a cumulative value of $\delta_i$'s as large as the one observed is obtained by chance. In this study, $2,500$ random realizations were performed per cluster, and clusters with $P(\Delta) \geq 0.1$ were considered to have passed the DS test.

The $P(\Delta)$ values, along with the $Z$-score and its confidence interval of the virialization indicator test described in the preceding section, are presented in Table~\ref{tab:resvirandsub} for the 14 clusters from the preliminary dataset of 37 whose $1\,R_{\mathrm{vir}}$ cores pass both tests. In this table, cluster 2236 is listed twice (A and B) because it hosts two nearly equally dominant BCGs (see next section). In this particular case, both BCGs must pass the indicator test, and since they do, we list the results separately.

\section{The maximally relaxed cluster sample}
\label{sec:results}

\subsection{Definition and properties}
\label{sec:resfinalclusters}

Of the 37 preselected evolved galaxy aggregations, 14 cluster regions were found to satisfy both the virialization and substructure criteria, and are therefore expected to host the most dynamically relaxed cores. This final set of cluster regions is hereafter referred to as the maximally relaxed (MR) cluster sample. The key properties of these 14 cluster regions are listed in Table~\ref{tab:finalclusters}.

\begin{table*}[]
    \centering
    \caption{Main properties of MR clusters.} 
    \label{tab:finalclusters}
    \begin{tabular}{cccccccccc}
        \toprule
        ID & Abell $\#^a$ & $N_{\mathrm{tot}}^b$ & $N_{\mathrm{clu}}$ & $z_{\mathrm{clu}}$ & $\mbox{R.A.}$[deg] & $\mbox{Dec.}$[deg] & $R_{\mathrm{vir}}$[Mpc] & $M_{\mathrm{vir}}$[$10^{14} \text{M}_{\odot}$] & $\sigma_{\mathrm{clu}}$[km\,s$^{-1}$] \\ 
        \midrule
        205 & A1691 & 205 & 89 & 0.073 & 197.8042 & 39.2410 & 2.7500 & 12.1095 & 877.61 \\
        262 & A1809 & 195 & 90 & 0.080 & 208.2826 & 5.1494 & 2.3297 & 7.3625 & 751.00 \\
        292 & A1904 & 239 & 107 & 0.072 & 215.5512 & 48.5482 & 2.5564 & 9.7271 & 814.73 \\
        571 & A1663 & 219 & 83 & 0.084 & 195.6563 & $-2.4993$ & 2.5500 & 9.6545 & 826.23 \\
        734 & A2028 & 153 & 65 & 0.077 & 227.3734 & 7.5756 & 2.2285 & 6.4442 & 715.59 \\
        804 & $-$ & 55 & 33 & 0.073 & 232.3093 & 52.8827 & 1.5800 & 2.2966 & 504.66 \\
        2236 & $-$ & 100 & 42 & 0.063 & 208.0137 & 46.3524 & 1.7392 & 3.0631 & 547.36 \\
        2316 & $-$ & 70 & 39 & 0.072 & 163.5068 & 54.8142 & 1.6252 & 2.4992 & 518.04 \\
        2496 & A1913 & 163 & 83 & 0.054 & 216.6968 & 16.6641 & 1.6612 & 2.6690 & 516.35 \\
        2620 & $-$ & 160 & 62 & 0.039 & 233.1474 & 4.6955 & 1.6657 & 2.6908 & 506.75 \\
        5183 & A1377 & 125 & 74 & 0.052 & 176.8621 & 55.7434 & 2.3043 & 7.1240 & 713.87 \\
        5567 & $-$ & 175 & 90 & 0.045 & 230.4534 & 7.6990 & 2.0760 & 5.2096 & 636.96 \\
        5905 & A1185 & 204 & 126 & 0.032 & 167.7492 & 28.6817 & 1.8806 & 3.8722 & 566.50 \\
        8062 & A1314 & 197 & 115 & 0.033 & 173.7296 & 49.0778 & 2.5326 & 9.4584 & 764.11 \\
                \midrule
        Mean$^c$ & $-$ & 161 & 78 & 0.061 & $-$ & $-$ & 2.1057 & 6.0129 & 661.42 \\
        \bottomrule
        \noalign{\smallskip}
    \end{tabular}
    \vspace{3pt}
    \parbox{\linewidth}{\footnotesize
    $^a$ Cluster number in the Abell catalog.\\
    $^b$ Total number of galaxies in the cluster region. Descriptions of the variables in the remaining columns are provided in the text. \\
    $^c$ Mean of the values in the columns, when applicable.
    }

\end{table*}

All clusters in the MR sample have virial radii that clearly exceed $1\,\mbox{Mpc}$, with a mean value of $2.1~\mbox{Mpc}$. The cluster virial masses range between $\sim 2.3\times 10^{14}~\mbox{M}_{\odot}$ and $\sim 1.2\times 10^{15}~\mbox{M}_{\odot}$, with an average of $\sim 6.0\times 10^{14}~\mbox{M}_{\odot}$. Cluster velocity dispersions span from approximately $500~\mbox{km}\,\mbox{s}^{-1}$ for the lower-mass clusters up to almost $880~\mbox{km}\,\mbox{s}^{-1}$ for the most massive systems, with a mean velocity dispersion of $\sim 660~\mbox{km}\,\mbox{s}^{-1}$. The total number of galaxies in this cluster sample is $2,260$, comprising $563$ S0s ($\sim 25\%$ of the total) and $1,697$ galaxies of other morphological types. Within the S0 population, 452 galaxies ($80\%$) are classified as S0--PS, 89 ($16\%$) as S0--AC, and 22 ($4\%$) as S0--TR. These proportions align with the findings of \citet{Tous2020} and \citet{Jimenez-Palau2022}, who reported that lenticular galaxies exhibiting signs of activity (i.e.,\ the AC and TR classes) tend to avoid high-density regions.

In Appendix \ref{app:plotsclusters}, we present plots showing the PPS of the MR cluster regions up to $3\,R_{\mathrm{vir}}$, as well as the corresponding sky distributions of their member galaxies along with visual aids to grasp how closely the spatial locations of the BCGs and the peaks of X-ray emission align with the clusters' central coordinates.

\subsection{BCG classification of MR clusters}
\label{BCGresults}

Table~\ref{tab:BCGs} lists the absolute magnitude, 2--1 and 3--1 magnitude gaps, and LOS velocities in the cluster rest frame of the first-ranked galaxies of the 14 MR cluster regions, together with their true BCG status. The last two columns of the table provide the projected distances from the BCGs to the center of the galaxy distribution, $d_{\mathrm{BCG,gal}}$, and to the peak of the extended X-ray emission, $d_{\mathrm{BCG,Xray}}$.

By examining the magnitude gaps between the three brightest galaxies, we find that in 8 MR clusters ($57\%$), the $\Delta{\cal{M}}_{2-1}$ gap exceeds 0.45 mag, classifying them as single-BCG systems. Three more clusters in the MR sample host a double BCG ($21\%$), while another three ---based on the classification criteria outlined in Sect.~\ref{sec:identifyBCGs}--- are categorized as non-BCG systems. 

A noteworthy case is the cluster region 2236, which hosts two nearly identical BCGs whose magnitudes only differ in 0.005 units. Given their comparable dominance, we applied the indicator test to both galaxies and found that both exhibit $Z$-scores consistent with the velocity distribution of the host cluster. Interestingly, the brightest BCG (labeled "A") is the most displaced BCG in the MR sample in terms of projected separation from both the galaxy distribution center and the X-ray peak (see Table~\ref{tab:BCGs} and the corresponding sky map in Appendix~\ref{app:plotsclusters}). In contrast, the sky position of its slightly fainter counterpart (B) is closely aligned with the X-ray emission center, although not so much with the average location of the cluster members.

Overall, the projected distances between BCGs and both the centers of the galaxy distributions and X-ray emission align with expectations, remaining relatively small compared to the virial radii of their host clusters, with only one of the two BCGs in the double-BCG cluster 2236 lying at a projected distance greater than 1 Mpc from the cluster’s X-ray center. Interestingly, in two of the three non-BCG galaxy aggregations where $\Delta{\cal{M}}_{2-1} < 0.45$, the projected distances between the BCGs and the X-ray peaks rank among the three largest in the MR sample, while the corresponding $P(\Delta)$ values are among the lowest that still pass the DS test. Both findings support the notion that the brightest galaxies in these systems are not truly dominant. Nonetheless, our sample size is too limited to conclude that these characteristics are definitive hallmarks of non-BCG relaxed clusters.

\section{The PPS distribution of S0s in the MR cluster sample}
\label{sec:PPSS0MRsample}

\subsection{For S0s divided into PCA-based spectral classes}
\label{sec:PPSS0MRsample-PCA}

The 2D distribution of S0 galaxies in the PPS diagram was examined for the MR cluster sample, extending to three virial radii. To ensure a consistent and precise characterization across all cluster regions, the 14 MR clusters were combined into a single stacked sample. This stacking was performed by normalizing each galaxy’s clustercentric radius by the virial radius of its host cluster, and by expressing its LOS velocity in units of the cluster’s velocity dispersion.

\begin{table*}[]
    \centering
    \caption{Properties of the BCGs in MR clusters.} 
    \label{tab:BCGs}
    \begin{tabular}{cccccccc}
        \toprule
        ID & $M_{\mathrm{BCG}}$[mag] & $\Delta{\cal{M}}_{2-1}$[mag] & $\Delta{\cal{M}}_{3-2}$[mag] & Status & $v_{\mathrm{BCG}}$[km\,s$^{-1}$] & $d_{\rm{BCG,gal}}$[Mpc] & $d_{\mathrm{BCG,Xray}}$[Mpc] \\
        \midrule
        205 & $-$23.169 & 0.525 & 0.790 & single & $-$35.99 & 0.128 & 0.167 \\
        262 & $-$23.213 & 0.980 & 0.348 & single & $-$176.69 & 0.036 & 0.133 \\
        292 & $-$23.024 & 0.900 & 0.342 & single & $-$208.25 & 0.132 & 0.390 \\
        571 & $-$21.836 & 0.169 & 0.077 & non-BCG & $-$129.18 & 0.483 & 0.479 \\
        734 & $-$22.867 & 1.231 & 0.234 & single & $+$72.36 & 0.431 & 0.385 \\
        804 & $-$23.052 & 1.581 & 0.254 & single & $-$159.77 & 0.106 & 0.296 \\
        2236A$^a$ & $-$22.585 & 0.005 & 0.995 & double & $-$153.01 & 0.706 & 1.104 \\
        2236B$^a$ & $-$22.579 & $-$0.005\ \ \ & 0.995 & double & $-$94.18 & 0.413 & 0.072 \\
        2316 & $-$22.107 & 0.621 & 0.084 & single & $-$23.24 & 0.644 & 0.238 \\
        2496 & $-$21.919 & 0.320 & 0.022 & non-BCG & $-$2.91 & 0.166 & 0.081 \\
        2620 & $-$22.249 & 1.242 & 0.041 & single & $+$9.27 & 0.060 & 0.286 \\
        5183 & $-$22.684 & 0.387 & 0.716 & double & $-$42.23 & 0.105 & 0.177 \\
        5567 & $-$22.492 & 1.058 & 0.368 & single & $-$166.14 & 0.055 & 0.003 \\
        5905 & $-$21.800 & 0.441 & 0.070 & non-BCG & $+$119.58 & 0.347 & 0.445 \\
        8062 & $-$22.271 & 0.552 & 0.623 & single & $+$56.09 & 0.062 & 0.037 \\ 
        \midrule
        Mean$^b$ & $-$22.519 & 0.715 & 0.355 & $-$ & $-$60.01 & 0.247 & 0.302 \\
        \bottomrule
        \noalign{\smallskip}
    \end{tabular}
    \vspace{3pt}
    \parbox{\linewidth}{\footnotesize
    $^a$ Two rows are included for cluster 2236 (A,B), one for each equally luminous BCG. \\
    $^b$ When applicable, the last row presents the mean of the values in the column.}
\end{table*}

\begin{figure*}[]
    \centering
    \includegraphics[width=0.9\textwidth]{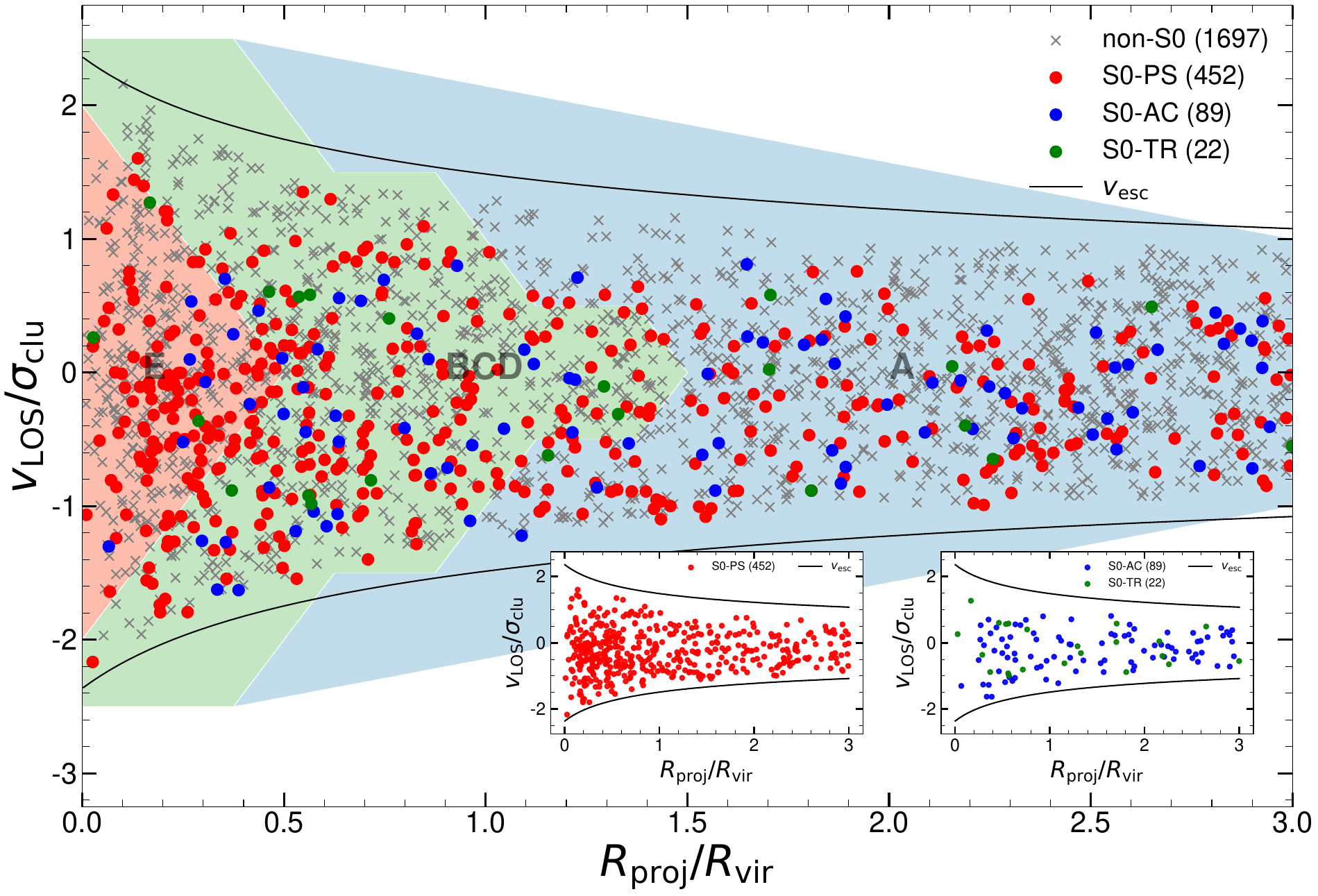}
    \caption{Projected phase-space diagram of all galaxies in the 14 clusters that constitute the most relaxed (MR) cluster sample. Non-S0 galaxies are shown as gray crosses, while S0 galaxies are represented by filled circles: Red for S0--PS, blue for S0--AC, and green for S0--TR. The projected radii and LOS velocities of all galaxies are normalized by the virial radius and velocity dispersion of their host clusters. The data are enclosed by the common caustics (black curves), which provide an estimate of the ensemble's escape velocity, while the colored shaded regions indicate the basic infall zones defined by \citet{Rhee2017} (see text for details). Two inset panels are included to emphasize the contrasting distributions of PS (left) and AC+TR lenticulars (right).}
    \label{fig:phasespace_allgals}
\end{figure*}

Fig.~\ref{fig:phasespace_allgals} presents the PPS distribution of the combined cluster regions, distinguishing between lenticular galaxies (filled circles) and other morphological types (gray crosses). Within the S0 population, galaxies are further identified according to their spectral activity class: S0--PS (red), S0--AC (blue), and S0--TR (green). This plot also includes the caustic lines derived assuming that the matter distribution of all the individual clusters in the stack follows a NFW density profile with a concentration parameter $C=6$. The two insets located at the bottom of the figure help to visualize the significant difference shown by the projected distributions of passive (PS, left) and active and mildly active (AC+TR\footnote{Due to their limited number, mildly active S0--TR galaxies will frequently be grouped with S0--AC galaxies in the present study.}, right) lenticulars along the radial coordinate (see below).

To quantify the observed differences, we compared the PPS distributions of the three spectral S0 subpopulations by performing 2D Kolmogorov-Smirnov (2D-KS) tests. For this analysis, we employed the publicly available \textsc{ndtest} code\footnote{Developed by Zhaozhou Li, \url{https://github.com/syrte/ndtest}.}. Following standard practice, we considered a two-sided $p$-value below $0.10$ as indicative of a statistically significant difference between the samples. As reported in Table~\ref{tab:KStest}, the 2D-KS test comparing the distributions of S0--PS and S0--AC yields a $p$-value of $0.0056$, confirming a statistically significant distinction between these two lenticular subpopulations. In contrast, the comparisons involving the S0--AC and S0--TR subsets, as well as between the S0--PS and S0--TR ones, result in larger $p$-values of $0.264$ and $0.267$, respectively. This aligns with the fact that S0--TR galaxies exhibit spectral characteristics intermediate between passive and active systems, as reflected in their position in the PC1-PC2 diagram. We note, however, that the statistical significance of results involving this latter spectral class should be interpreted with caution due to the limited size of this subset. We also used the 2D-KS test to compare the PPS distributions of the entire S0 population with the rest of the cluster galaxies, obtaining a two-sided $p$-value of only $0.0002$ indicative of a high statistical inconsistency between both distributions. This suggests that S0 galaxies are not merely a random subset of the general cluster population, but instead follow a distinct evolutionary trajectory, likely shaped by differences in their formation history and the environmental mechanisms influencing galaxy evolution.

\begin{table}[]
    \caption{Results of KS tests for the various spectral and $\Delta\mbox{MS}$ classes of S0s.}
    \label{tab:KStest}
    \begin{tabular}{ccccc}
    \toprule
    \multicolumn{2}{c}{Comparison$^a$} & \multicolumn{3}{c}{$p$-values$^b$} \\
    \midrule
    Sample 1 & Sample 2 & PPS & $R_{\mathrm{proj}}/R_{\mathrm{vir}}$ & $v_{\mathrm{los}}/\sigma_{\mathrm{clu}}$ \\ 
    \midrule
    S0--PS & S0--AC & 0.0056 & 0.0023 & 0.6023 \\
    S0--PS & S0--TR & 0.2670 & 0.6132 & 0.5288 \\
    S0--AC & S0--TR & 0.2640 & 0.4509 & 0.5317 \\
    \midrule
    S0--QS & S0--MS & 0.0046 & 0.0004 & 0.9309 \\
    S0--QS & S0--GV & 0.0242 & 0.0061 & 0.8592 \\
    S0--MS & S0--GV & 0.4130 & 0.3293 & 0.9639 \\
    \midrule
    S0 & other types & 0.0002 & 0.0003 & 0.0103  \\
    \bottomrule
    \noalign{\smallskip}
    \end{tabular}
    \vspace{3pt}
    \parbox{\linewidth}{\footnotesize
    $^a$ Samples included in the comparison.\\
    $^b$ Two two-sided $p$-values from the 2D-KS test on the PPS (3rd column) and from the classical 1D-KS test on each one the two dimensions of the PPS (4th and 5th columns).}
\end{table}

\begin{figure}[]
    \centering
    \includegraphics[width=\columnwidth]{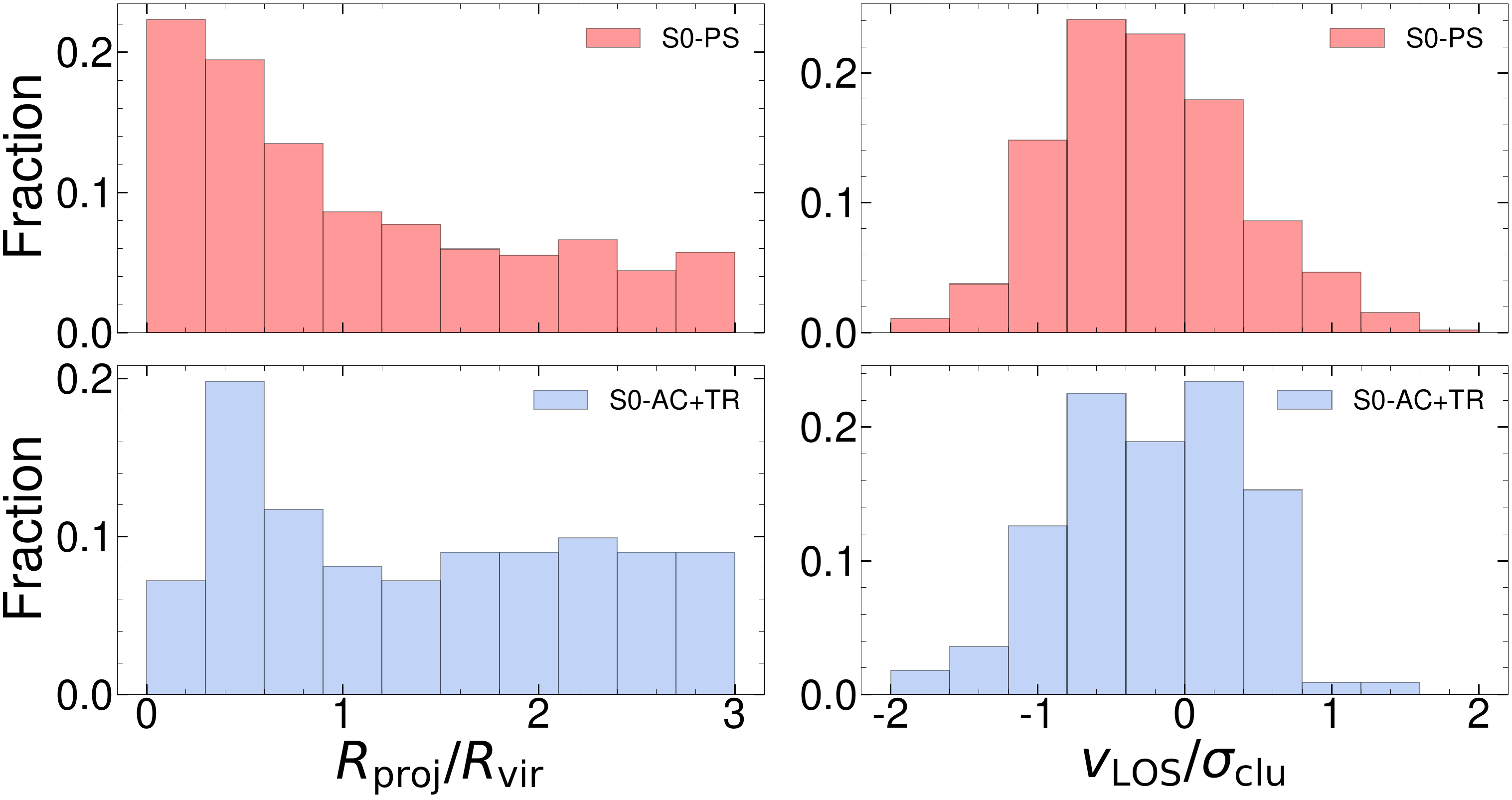}
    \caption{Histograms of dimensionless projected clustercentric distance (left) and LOS peculiar velocity (right) for the passive (PS, top, red) and non-passive (AC+TR combined, bottom, light-blue) spectral classes of S0 galaxies in the MR cluster sample. A test of normality \citep{dagostino1971normailty} yields $p$-values of 0.3243 and 0.3785 for S0--PS and S0--AC+TR, respectively.}
    \label{fig:S0s_kde}
\end{figure}

Further graphical evidence of the distinct distributions of passive (PS) and non-passive (AC+TR) S0 galaxies in the PPS of the MR cluster ensemble is provided in Fig.~\ref{fig:S0s_kde}. This figure presents histograms of the marginal distributions of the dimensionless projected clustercentric distances (left panels) and LOS peculiar velocities (right panels) of the two spectral subpopulations. The clustercentric distance distributions reveal that S0--PS galaxies are both more centrally concentrated and more numerous at small radii, increasing monotonically toward the cluster center, while their AC+TR counterparts peak around $0.5\,R_{\mathrm{vir}}$ before declining sharply in the innermost radial bin. Beyond $\sim 1.5\,R_{\mathrm{vir}}$, both subpopulations maintain nearly constant radial abundances, extending to the outer edge of the infall region at  $3\,R_{\mathrm{vir}}$. Despite this, the pronounced differences in the inner region lead a classical two-sample 1D-KS test to reject the null hypothesis that the S0--PS and S0--AC distributions are the same with a confidence level of $0.0023$. In contrast, comparisons involving the S0–TR subset yield $p$-values well above 0.10; see Table~\ref{tab:KStest}. For their part, the differences in the distributions of LOS velocity dispersion are much less pronounced, with all three spectral S0 subpopulations exhibiting roughly bell-shaped distributions and statistically insignificant differences reflected in $p$ values above $0.5$. These findings reveal that the discrepancy in the PPS distributions of active and passive S0s is driven primarily by spatial segregation. Following the approach used with the 2D-KS test, we applied a two-sided 1D-KS test to compare the radial and LOS velocity distributions of the entire S0 population with those of non-S0 cluster galaxies. In both dimensions, we obtained $p$-values $\lesssim 0.01$, reinforcing the notion that these two ensembles  have completely distinct origins.

Cluster galaxies were also studied based on their location in phase space, which serves as a proxy for the time elapsed since they became part of the cluster region. Following the methodology of \citet{Rhee2017}, the PPS diagram of the cluster ensemble shown in Fig.~\ref{fig:phasespace_allgals} was divided into distinct regions corresponding to different typical infall times (see their Fig.~6). In our plot the five original domains defined by these authors were consolidated into three by merging the intermediate regions B, C, and D into a single zone labeled BCD. This adjustment was made to ensure that each infall region contains a statistically meaningful number of both TR and AC lenticulars. According to \citeauthor{Rhee2017}, galaxies in region A are recent arrivals, likely undergoing their first infall, while those in region BCD ---most of which reside within $1\,R_{\mathrm{vir}}$--- are intermediate infallers accreted at earlier times. Meanwhile, galaxies in region E represent ancient infallers and, hence, the oldest cluster members. 

\begin{figure}[]
     \centering
     \includegraphics[width=\columnwidth]{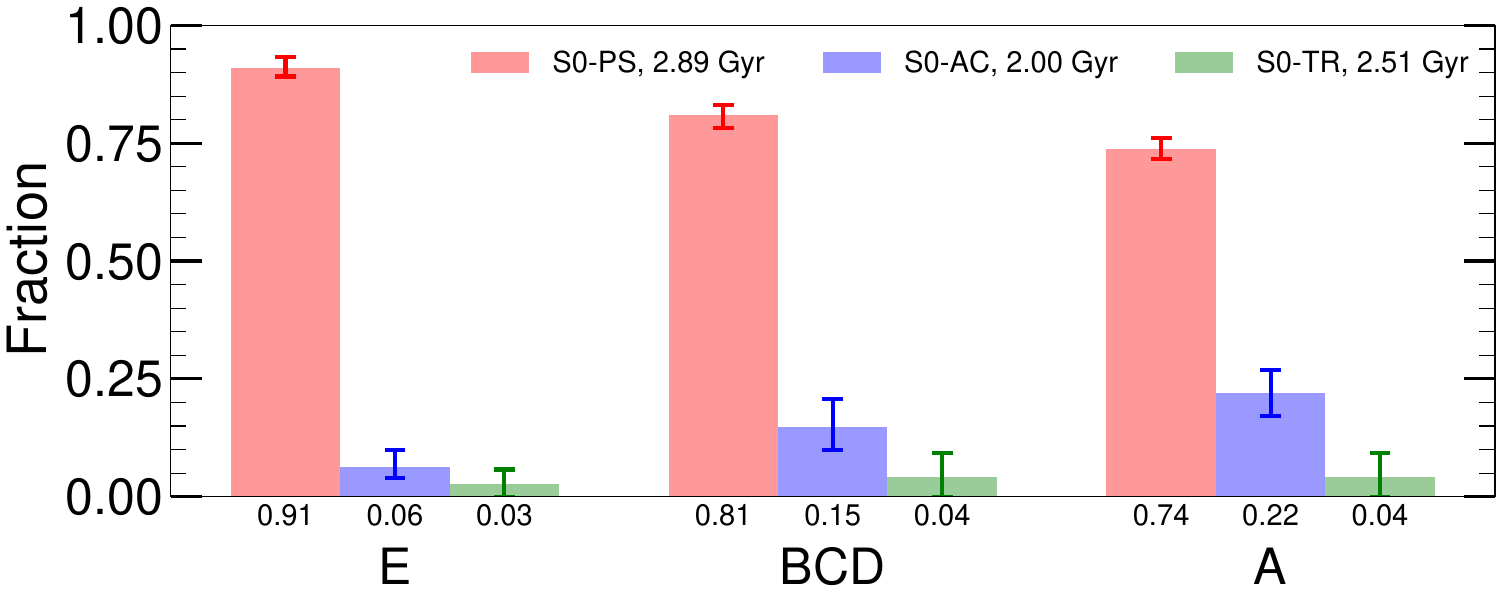}
     \caption{Fractions of the various S0 spectral classes across the PPS infall regions defined in Fig.~\ref{fig:phasespace_allgals}. Red bars represent S0--PS, blue bars S0--AC, and green bars S0--TR galaxies, with the numbers below the bars indicating their fractional abundances. The numbers following each label denote the average time since cluster infall for each S0 class, calculated by weighting their relative abundance in each infall region by the mean time since infall (see text). The $1\,\sigma$ error bars were derived from 1000 bootstrap resamples with replacement of the original distributions in each infall region.}
     \label{fig:S0s_regfracs}
\end{figure}

In Fig.~\ref{fig:S0s_regfracs}, we present the fractional distribution of the three spectral S0 classes across the three main infall dominions just defined in the PPS diagram. This figure shows that the S0--PS galaxies are the most prevalent in all infall regimes and have a relative abundance that systematically increases toward the cluster core. In contrast, S0--AC lenticulars show a steady decline in their fraction from the outskirts toward the center. Meanwhile, S0--TR galaxies display a neutral distribution in the outer A and BCD regions, plus an absolute drop of $1\%$ (but $25\%$ in relative terms) in the innermost E zone, which is consistent with their intermediate spectral characteristics. However, we note again that the small size of this latter subset introduces substantial statistical uncertainty, which limits the robustness of any conclusions drawn from this group. 

We also estimated the typical mean time since first infall into the cluster region for each spectral class of S0 galaxies. This was calculated by weighting the relative abundance of each spectral class in a given infall region by the mean time since infall assigned to that region in \citet{Oh2019}: 5.82 Gyr for region E (ancient infallers), 3.52 Gyr for region BCD (intermediate infallers), and 0.44 Gyr for region A (recent infallers). Using this method, we derived mean infall times of 2.89 Gyr for S0--PS, 2.51 Gyr for S0--TR, and 2.00 Gyr for S0--AC. Of greater interest is the fact that there is a clear difference of $\sim 0.9$~Gyr in the membership span between the passive and active S0 populations. This discrepancy aligns with their differing radial concentrations in the PPS, indicating that S0--PS galaxies have resided in the cluster environment for significantly ($\sim 45\%$) longer and that, consequently, have had more time to virialize.

\subsection{For S0s divided into SSFR-based classes}
\label{sec:PPSS0MRsample-DMS}

As shown in \citet{Jimenez-Palau2022} and \citeauthor{Tous2024} (\citeyear{Tous2024, Tous2025}), in the PCA-based taxonomy ---which is sensitive to any type of nebular ionization source reflected in the optical spectra of galaxies--- the S0--AC spectral class is predominantly composed of SF S0s, while the S0–PS group mainly consists of quiescent galaxies. This strong correlation between our PCA classification and the star-formation status of present-day S0 galaxies is expected to be even more pronounced in clusters, where active galactic nuclei are rarely found \citep[e.g.,][]{Marziani2017}. As a result, a more explicit means of assessing the effects of the cluster environment can be employed. Instead of identifying these effects generically through their impact on optical spectra, a classification scheme can be adopted in which environmental influence is quantified specifically through star formation activity, thereby minimizing the confounding effects of nuclear activity and alternative ionization sources.

To achieve this, we introduced the parameter $\Delta\mbox{MS}$, which measures the SSFR of cluster S0s relative to the MS ridge of SF galaxies \citep{Renzini2015}. This metric is computed following \citet{Jimenez-Palau2022} as:
\begin{equation}
    \Delta\mbox{MS} = \log \left[ \frac{\text{SFR}}{\text{M}_\odot ~ \text{yr}^{-1}} \right] - 0.76 \log \left[ \frac{M_\star}{\text{M}_\odot} \right]  + 7.6\;,
\end{equation}
where the information about the star-formation rates (SFR) and stellar masses ($M_\star$) of the galaxies was retrieved from \citet{Salim2018}, who derived these properties by fitting the observed spectral energy distributions of these objects across the UV, optical, and IR electromagnetic domains. Following \citet{Tous2024}, we then classified S0 galaxies as follows: Main Sequence lenticulars (S0--MS) have $\Delta\mbox{MS} \ge -0.5$, Green Valley galaxies (S0--GV) verify $-0.5 > \Delta\mbox{MS} > -1.1$, and Quenched objects (S0--QS) are characterized by $\Delta\mbox{MS} \le -1.1$. Among the 563 S0s in the MR sample, 83 ($15\%$) are classified as S0--MS, 98 ($17\%$) as S0--GV, and 382 ($68\%$) as S0--QS. 

\begin{figure}[]
    \centering
    \includegraphics[width=\columnwidth]{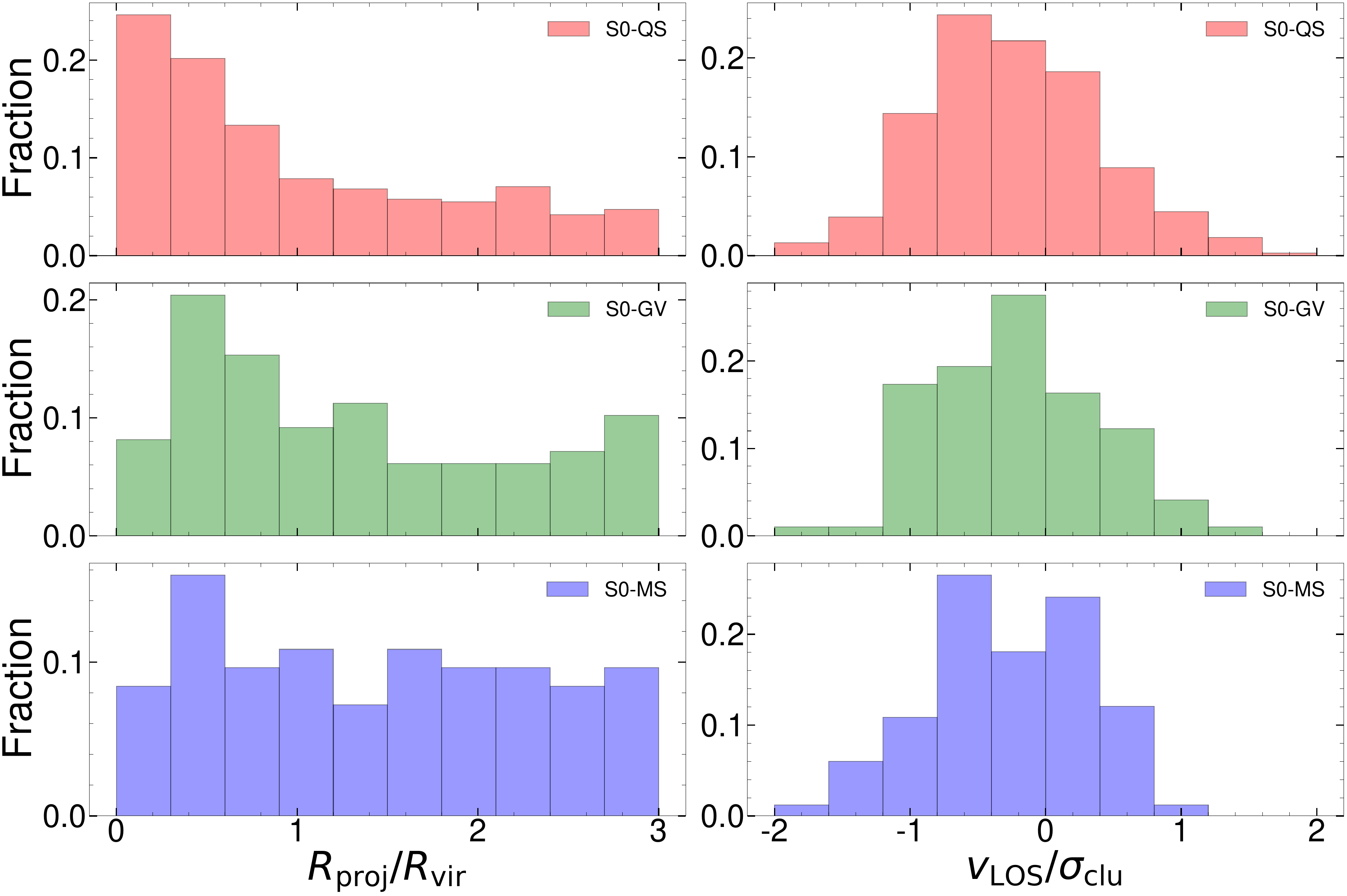}
    \caption{Same as Fig.~\ref{fig:S0s_kde} but for the $\Delta\mbox{MS}$ classes of lenticular galaxies in the MR cluster sample. The top red histograms represent S0--QS galaxies, the middle green histograms correspond to S0--GV galaxies, and the bottom blue histograms depict S0--MS galaxies. A test of normality \citep{dagostino1971normailty} yields $p$-values of 0.4413, 0.4708, 0.3168 for S0--QS, S0--GV, and S0--MS, respectively.}
    \label{fig:S0s_kde_DMS}
\end{figure}

\begin{figure}[]
    \centering
    \includegraphics[width=\columnwidth]{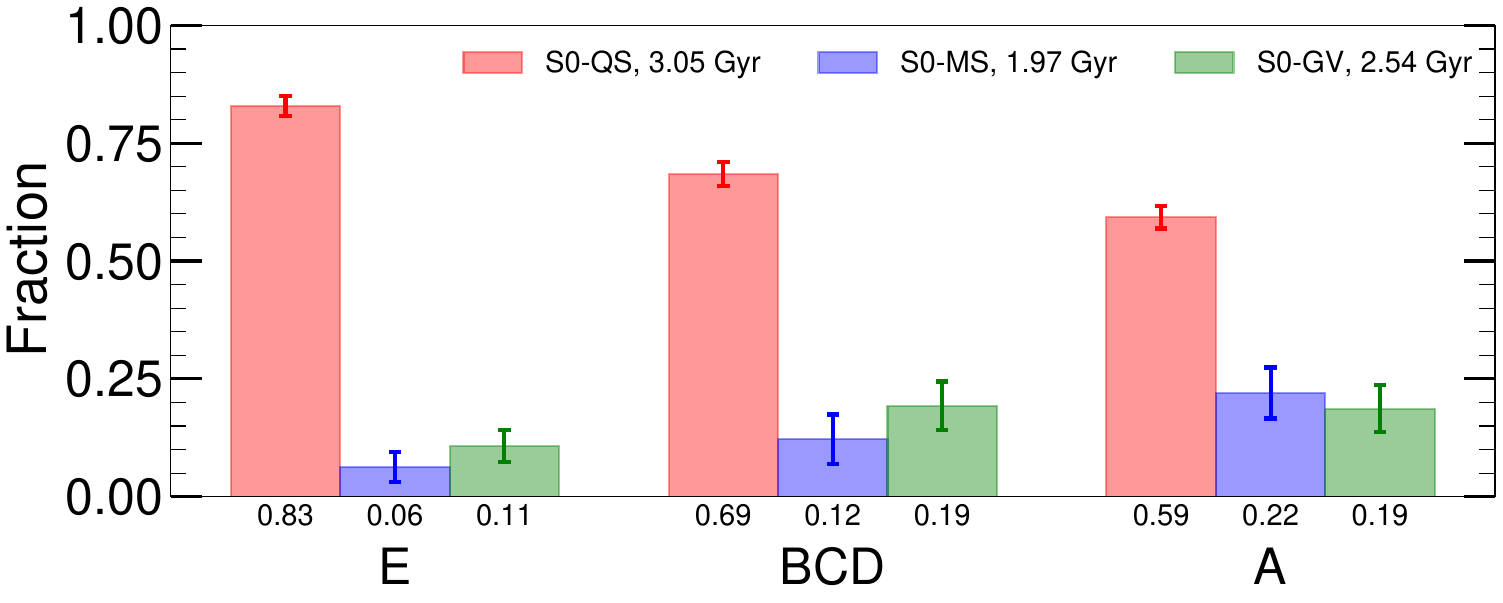}
    \caption{Same as Fig.~\ref{fig:S0s_regfracs} but for the fractions of $\Delta\mbox{MS}$-based classes. The red bars denote S0--QS, blue bars S0--MS, and green bars S0-GV galaxies.}
    \label{fig:DMS_regfracs}
\end{figure}

Next, we followed the same steps adopted previously to analyze the distributions of PCA-based classes in the PPS of the stack of MR clusters. First, we investigated differences in the 2D PPS distributions among the newly defined S0 classes, both visually through Fig.~\ref{fig:phasespace_allgals_DMS} in Appendix~\ref{extra_material}, which reaffirms a negative correlation between star formation activity and clustercentric distance, and quantitatively through the 2D-KS test, which yields results that conform reasonably well with those derived for the commensurate spectral classes. In particular, as shown in Table~\ref{tab:KStest}, the comparison between the S0--QS and S0--MS subsets results in a $p$-value of 0.0046, confirming that these two extreme S0 subpopulations are statistically distinct. In contrast, the MS and GV subsets yield a $p$-value of 0.413, indicating a relatively high degree of similarity. The only result that draws the attention is the comparison between the QS and GV classes, which now yields a $p$-value of 0.0242, suggesting a statistically significant difference not observed between the corresponding PS and TR PCA classes. However, this discordance in the outcomes may be simply explained by the notably small fraction of S0--TR, galaxies ($4\%$), which could limit the statistical power of the original comparison.

This alignment between PCA and $\Delta\mbox{MS}$-based classifications of S0s is additionally reinforced in Fig.~\ref{fig:S0s_kde_DMS}, which displays histograms of the marginal distributions of the dimensionless projected clustercentric distance and LOS peculiar velocity for the various $\Delta\mbox{MS}$ classes. The shapes of these histograms closely resemble those shown in Fig.~\ref{fig:S0s_kde}, with the two-sample 1D-KS test now revealing significant differences in the distributions of projected radii between the S0--QS and both the S0--MS and S0--GV subsets  ---which, as before, primarily arise in the innermost cluster regions---, and full consistency in the corresponding velocity distributions (see Table~\ref{tab:KStest}). The fractional distributions of $\Delta\mbox{MS}$ classes across the PPS infall regions (Fig.~\ref{fig:DMS_regfracs}) further corroborate these results, closely mirroring the trends observed in the analogous spectral classes (Fig.~\ref{fig:S0s_regfracs}). In particular, there is a steady increase in the fraction of S0--QS systems toward the cluster core, accompanied by an inverse trend for the S0--MS class of SF systems\footnote{This agreement confirms that the presence of galaxies with ionization sources unrelated to star formation in the PCA classification has a negligible impact on the overall results.}. For their part, the S0--GV galaxies, similar to the S0--TR galaxies, display a nearly constant abundance in the two outer infall regions, followed by a decline in the central region which is also the most densely populated. Additionally, we note that while the absolute variations in the fractions of active or SF galaxies may appear modest due to the overwhelming presence of passive or quiescent objects in all infall zones, the relative changes are substantial. This reinforces the notion that the former represent a subpopulation of S0s that has entered the cluster environment more recently. 

Likewise, the inferred typical mean times since first cluster infall are also in line with all these findings. Using the same methodology as before, we derived mean infall times of $3.05$ Gyr for S0--QS, $2.54$ Gyr for S0--GV, and $1.97$ Gyr for S0--MS. These results reaffirm that the most quiescent (passive) S0 galaxies have, on average, resided in the cluster environment for approximately $1$ Gyr longer than their most SF (active) counterparts, highlighting a substantial difference in their accretion timescales.

\section{Radial VDLOS and SSFR profiles of S0s in the MR
cluster sample}
\label{sec:profiles}

In this section, we construct radial VDLOS and SSFR profiles as functions of the projected clustercentric radius for the stack of MR clusters. This approach provides a continuous characterization of the PPS distribution of the different S0 subpopulations, offering a more detailed perspective than the coarse binning across infall zones adopted in the previous section. In addition, by co-adding the clusters, we achieve a more robust determination of profile behavior at large projected distances, where individual cluster samples become increasingly sparse. The stacking also helps mitigate potential distortions arising from undetected substructures and deviations from spherical symmetry.

\subsection{Radial VDLOS profiles}
\label{sec:profiles-VDP}
The radial VDLOS profiles for different classes of S0 galaxies allow us to examine how their classification correlates with their kinematics in the cluster environment.

These profiles were computed using the method outlined in \cite{Bilton2018}. In brief, the approach is based on Gaussian kernel density estimation \citep[e.g.,][]{silverman2018density}, but rather than applying the Gaussian kernel directly to the measurements, it is applied to their radial coordinates. This determines the weighting of each measurement’s contribution at a given projected radius $R$ as follows: 
\begin{equation}
\label{eq:Gkernel}
    w_i(R) = \frac{1}{\sqrt{2\pi}\sigma} \exp\left[- \frac{(R - R_i)^2}{2 \sigma_w^2}\right]\;,
\end{equation}
where $\sigma_w$ is the kernel bandwidth, which sets the strength of the smoothing, and $R_i$ the projected clustercentric radius of the galaxy providing the measurement. These weights were then used to calculate the radial VDLOS profile from the expression 
\begin{equation}
\label{eq:profiles}
    \sigma_{\mathrm{LOS}}(R) = \sqrt{\frac{\sum_i{w_i(R)(x_i - \bar{x})^2}}{\sum_i{w_i(R)}}}\;,
\end{equation}
with $x_i$, the observation of interest for each galaxy in the sample, represented here by the LOS velocity in the cluster rest frame (eq.~[\ref{eq:vlos}]), and $\bar{x}$ the sample mean of the observations, which in this particular case corresponds to the recession velocity of the cluster, $v_{\mathrm{clu}}$. After some experimentation with different values, the smoothing window was chosen to be $0.4\,R_{\mathrm{vir}}$ units, providing sufficient smoothing for an effective limitation of spurious data artifacts but not too large to obscure the basic underlying structure.

To provide reference points for comparison with the radial VDLOS profiles of S0s, we also constructed VDLOS profiles for various subsets of non-lenticular galaxies within the MR sample. Specifically, we derived profiles for 232 elliptical (E) and 782 spiral (S) galaxies subdivided into 657 early spirals (Sa+Sb) and 125 late spirals (Sc+Sd), with all morphologies taken from \citet{DomínguezSánchez2018}, as we did for the S0s. Additionally, we included a sample of 410 galaxies classified as SF in the classical BPT diagram \citep{Baldwin+1981} by the Portsmouth stellar kinematics and emission-line flux measurements from \citet{Thomas+2013}. The S0 population itself was divided into distinct subsets: 111 non-passive S0--AC+TR galaxies and 452 passive S0--PS objects, according to their PCA spectral classification, as well as in 83 S0--MS, 98 S0--GV, and 382 S0--QS galaxies according to the $\Delta\mbox{MS}$ scheme. All these sample sizes correspond to galaxies within $3\,R_{\mathrm{vir}}$; however, all galaxies extending out to $4\,R_{\mathrm{vir}}$ were used for radial profile calculations to minimize edge effects caused by data truncation in the outer $3\,R_{\mathrm{vir}}$-boundary of the cluster regions.

\begin{figure}[]
    \centering
    \includegraphics[width=0.95\columnwidth]{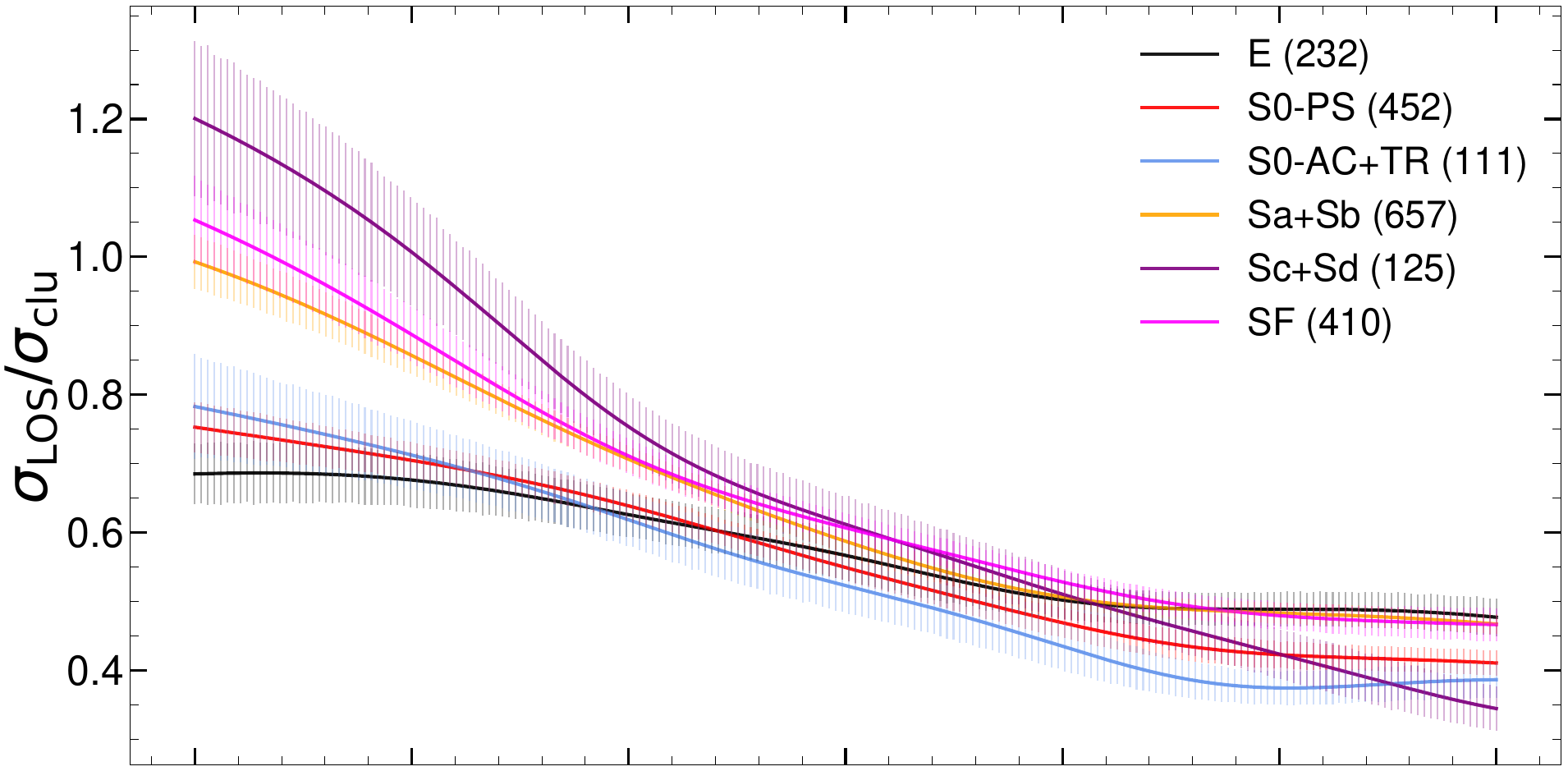}
    \includegraphics[width=0.95\columnwidth]{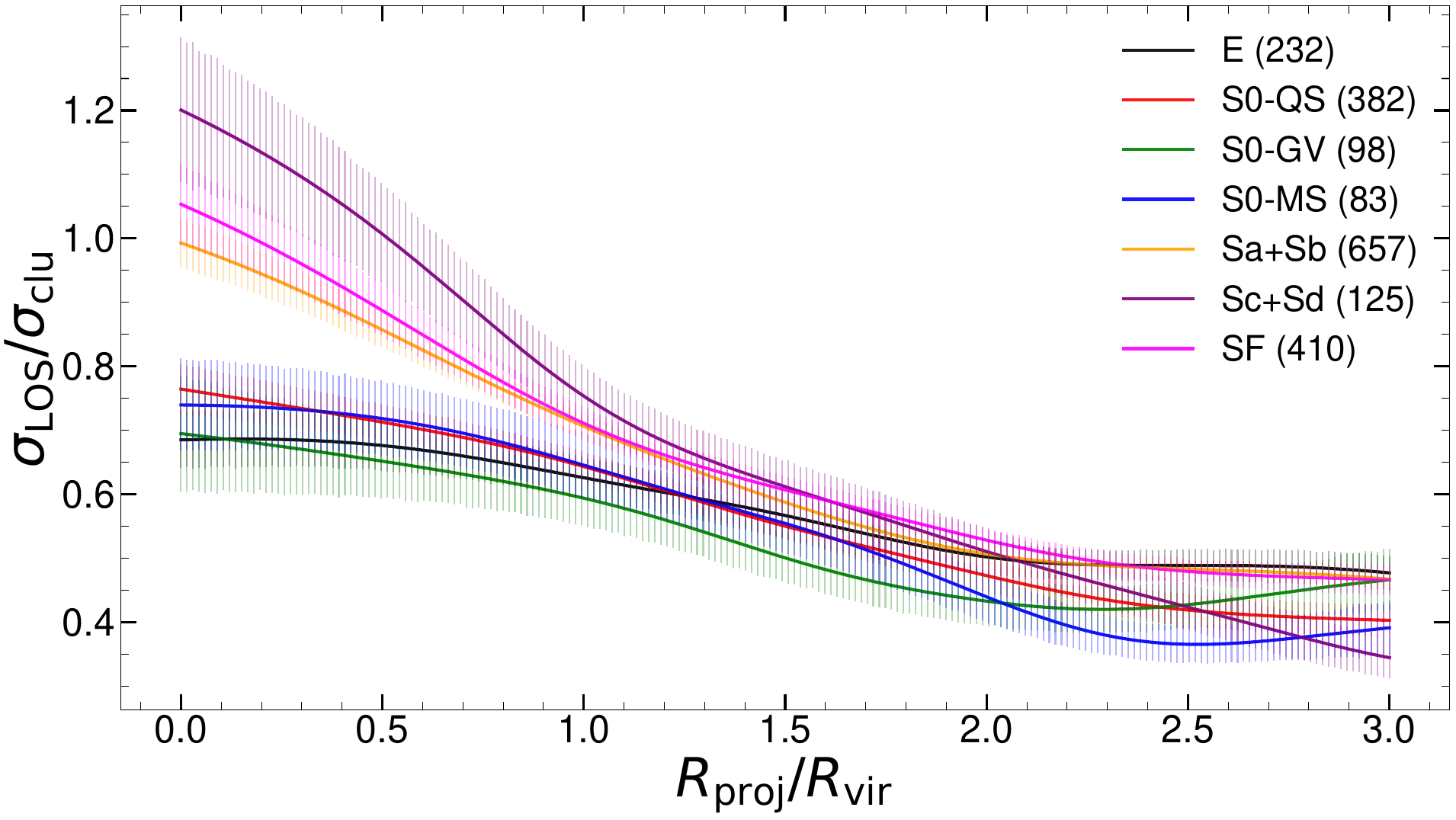}
    \caption{Top panel: Radial VDLOS profiles for PCA-classified lenticular galaxies: S0--AC+TR (light blue) and S0--PS (red), smoothed with a Gaussian kernel of bandwidth $0.4,R_{\mathrm{vir}}$. Bottom panel: Same as the top row, but using the $\Delta\mbox{MS}$-based classification: S0--MS (blue), S0--GV (green), and S0--QS (red). Both panels also include profiles for elliptical (E, black), early-type spiral (Sa+Sb, orange), late-type spiral (Sc+Sd, purple), and star-forming (SF, magenta) galaxies. Clustercentric distances are normalized by the virial radius of each host cluster. Profiles are computed using data out to $4,R_{\mathrm{vir}}$, although the sample sizes shown in the legends correspond to galaxies within $3,R_{\mathrm{vir}}$ (see text). Shaded regions represent 1$\sigma$ confidence intervals from 1000 bootstrap resamples.}
    \label{fig:vel_profs}
\end{figure}

As shown in Figure~\ref{fig:vel_profs}, the VDLOS profiles remain relatively flat and similar within the uncertainties across all galaxy classes in the outermost cluster region. A noticeable increase in velocity dispersion emerges around $2\,R_{\mathrm{vir}}$, progressing comparably across all classes until reaching $\sim 1\,R_{\mathrm{vir}}$, where the transition into the virialized cluster core occurs. Within this radius, the profiles split into two distinct trends. SF galaxies and the two subsets of spiral galaxies exhibit a pronounced inward increase in velocity dispersion, whereas the subsets of early-type galaxies, including all S0s regardless of their level of activity, show overall peculiar velocities that tend to plateau. Among early-type objects, E galaxies exhibit the weakest inward increase in velocity dispersion, managing to maintain a roughly constant VDLOS within $\sim 0.5\,R_{\mathrm{vir}}$. For their part, the subpopulations of the most active S0s (AC+TR, MS and GV) tend to display very mild inward rises in their systemic velocities. Nevertheless, all these profiles are statistically indistinguishable with one another within the limits of uncertainty.

Stellar dynamics shows that dynamically relaxed systems dominated by intrinsically radial orbits naturally produce monotonically declining VDLOS profiles with increasing distance to the center ---the more radially elongated the orbits, the steeper the decline. Accordingly, the observed behavior of these profiles in the core of the stack of MR clusters indicate that Sc+Sd galaxies follow the most elongated orbits, whereas those of SF and Sa+Sb galaxies display progressively less pronounced radial anisotropies. On the other hand, the relatively constant velocity dispersion profiles with projected radius exhibited by the E and all S0 subpopulations are characteristic ---though not exclusive--- of isotropic velocity distributions, a hallmark of virialized systems. Nevertheless, it is striking that the clear differences in projected clustercentric distances and average infall times between passive and active S0 classes inferred in Sect.~\ref{sec:PPSS0MRsample} barely translate to differences in the shapes of their inner VDLOS profiles. This suggests that even the most active S0s must be in a relatively advanced stage of dynamical relaxation within the cluster environment. In contrast, the significantly steeper profiles of the Sc+Sd or SF subsets support the idea that most of these galaxies are experiencing their first infall and/or spending more time in the cluster's outskirts, likely near the apocenters of their radially elongated orbits (see also Sect.~\ref{sec:profiles-SSFR}). Conversely, the observed similarity in the velocity profiles across all galaxy subsets beyond the virial radius suggests that not only late-type disks but also a fraction of lenticular galaxies, including their most passive representatives, acquire their morphology outside cluster cores and are subsequently accreted via secondary infall.

\begin{figure*}[]
    \centering
    \includegraphics[width=\textwidth]{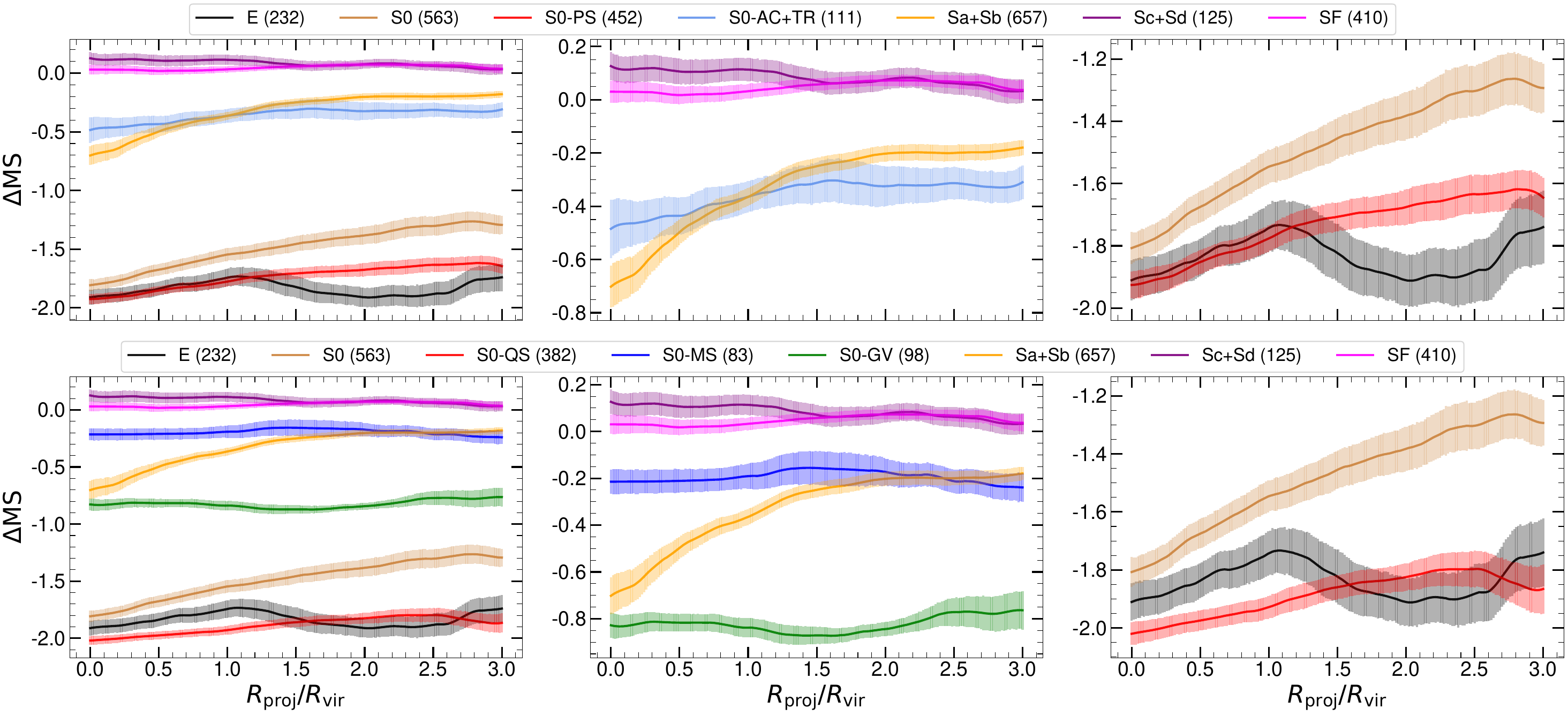}
    \caption{Top row: Radial $\Delta$MS profiles for PCA-classified lenticular galaxies: S0--AC+TR (light blue) and S0--PS (red), smoothed with a Gaussian kernel of bandwidth $0.4,R_{\mathrm{vir}}$. Bottom row: Same as the top row, but using the $\Delta\mbox{MS}$-based classification: S0--MS (blue), S0--GV (green), and S0--QS (red). Each row also includes profiles for elliptical (E, black), early-type spiral (Sa+Sb, orange), late-type spiral (Sc+Sd, purple), star-forming (SF, magenta), and the full S0 population (brown). Columns show: left, all galaxy subsets; center, only those with evidence of star formation; right, the most quiescent ones. Clustercentric distances are normalized by the virial radius of each host cluster. Profiles are computed using data out to $4,R_{\mathrm{vir}}$, although the sample sizes shown in the legends correspond to galaxies within $3,R_{\mathrm{vir}}$ (see text). Shaded regions represent 1$\sigma$ confidence intervals from 1000 bootstrap resamples.}
    \label{fig:ssfr_profs}
\end{figure*}

\subsection{Radial SSFR profiles}
\label{sec:profiles-SSFR}

Here, we analyze how the star formation activity of S0 galaxies, quantified by both their $\Delta$MS and SSFR values, depends on projected clustercentric distance, a proxy for environmental density. By comparing the trends of these parameters from both different S0 classes and other galaxy morphologies across the same environmental gradient, we aim to better understand how intrinsic properties and external influences combine to shape the evolution of the lenticular population.

To calculate the radial profiles of star formation activity, we modified the methodology used previously for constructing radial velocity profiles. Unlike velocity measurements, the star formation activity indicators ($\Delta\mbox{MS}$ and SSFR) do not follow a symmetric, bell-shaped distribution. Consequently, we determined the central tendency at each projected radius $R$ using the trimean (TRI), a computationally simple estimator that uses ordered statistics to improve somewhat upon the reliance of the median on one or two data points. The TRI estimator is defined as an average of the sample median, $Q_2$, and the lower, $Q_1$, and upper, $Q_3$, quartiles \citep[cf.][]{Beers1990}:
\begin{equation}
\label{eq:TRI}
C_{\mathrm{TRI}} = \frac{1}{4}(Q_1+2Q_2+Q_3)\;.
\end{equation}
In practice, to calculate $C_{\mathrm{TRI}}(R)$, we first applied the Gaussian kernel described in Eq.~(\ref{eq:Gkernel}) with $\sigma_w=0.4\,R_{\mathrm{vir}}$ to each measurement\footnote{As with the VDLOS profiles, we use data up to $4\,R_{\mathrm{vir}}$ to minimize edge effects at large radii.} to determine its weight $w_i(R)$. These weights were then normalized by the smallest weight and the ratios approximated to the nearest integer. The sum, $N^* = \sum_i w^*_i$, of these integer-normalized weights, $w^*_i$, was used to construct a synthetic sample of size $N^*$, in which each measurement is repeated $w^*_i$ times. Finally, the data of the synthetic sample was sorted in ascending order and the trimean computed using the formula above. The $1\,\sigma$ confidence intervals for this statistic were estimated by repeating this process 1000 times. 

The TRI statistic, however, yields noisy profiles with short-term spurious fluctuations due to its reliance on only a few data points. To suppress this high-frequency noise while preserving the underlying signal, we applied a Savitzky-Golay (SG) filter \citep{Savitzky-Golay64}. This filter smooths the data by fitting a low-degree polynomial via least squares to successive subsets of adjacent points within a moving window. For our star formation activity profiles, we adopted a 25-point window and a cubic polynomial smoothing function.

We applied these procedures to the same galaxy subsets defined in the previous section to provide a detailed assessment of how star formation activity varies within the cluster environment as a function of galaxy morphology, with a particular focus on the behavior of the different lenticular classes defined by both the PCA and $\Delta\mbox{MS}$ schemes. Our analysis incorporates two distinct measures of SSFR: its logarithm and the offset $\Delta\mbox{MS}$ from the MS. Figure~\ref{fig:ssfr_profs} presents a comparison of the radial $\Delta\mbox{MS}$ profiles, with the top row displaying three panels that contrast these profiles for PCA-based S0 classes against those of other galaxy types, while the bottom row uses the $\Delta\mbox{MS}$-based S0 classification for the comparison. The results of a similar analysis, but using the somewhat noisier logarithm of SSFR as the measure of star formation activity, are displayed in Fig. \ref{fig:ssfr_profs_DMS} in Appendix \ref{extra_material}.

The first observation from these plots is the consistency shown by the all S0 profiles in four sets of panels, highlighting the robustness of our results and demonstrating their relative independence from both the lenticular classification scheme and the chosen star formation activity metric. The top panels, which provide an overview of the radial $\Delta\mbox{MS}$ and SSFR profiles for all galaxy subsets, clearly reveal a pronounced gap of approximately corresponding to 1.5 orders of magnitude in SSFR that separates the most active (AC+TR and MS) S0 subpopulations from the most passive ones (PS and QS). In fact, as shown in the first and middle columns of Figs.~\ref{fig:ssfr_profs} and \ref{fig:ssfr_profs_DMS}, the most active S0 classes exhibit star-formation levels comparable to those of Sa+Sb galaxies, with the S0--AC subset in particular displaying a steady decline from $\sim 1.5\,R_{\mathrm{vir}}$ to the cluster center, much like the latter. In contrast, galaxies with the highest star-formation levels, such as late-type Sc+Sd spirals and SF BPT class members, display largely flat radial SSFR profiles, with the Sc+Sd galaxies even showing a modest SSFR increase within $\sim 1.5,R_{\mathrm{vir}}$, suggesting insignificant environmental influence. Notably, this apparent insensitivity to environmental effects also extends to the S0--MS and GV classes, which also maintain nearly constant star-formation levels ---significantly lower in the latter case--- throughout the entire cluster region. In Sect.~\ref{sec:discussion}, we offer a possible explanation for this remarkable behavior.

On the other hand, the S0--PS and QS classes show, as expected, the lowest levels of star formation, as low indeed as those shown by the E galaxy population (see the first and last columns of Figs.~\ref{fig:ssfr_profs} and \ref{fig:ssfr_profs_DMS}). Additionally, the radial $\Delta\mbox{MS}$ and SSFR distributions of these passive S0s, similar to their AC+TR counterparts, exhibit obvious signs of environmental influence, manifested, in this case, as a gradual decrease in SSFR starting from the outskirts of the MR cluster stack. In contrast, while the profiles for E galaxies also begin to decline at the periphery, they show a noticeable upturn around $2\,R_{\mathrm{vir}}$ that persists until $\sim 1\,R_{\mathrm{vir}}$, where their minimal activity levels temporarily match or exceed those of quiescent S0s before resuming their decline within the virial region.

Next, we provide a detailed discussion of the results from our analysis, including the trends observed in the profiles presented in this section, and their implications for the evolutionary history of the lenticular galaxy population.

\section{Discussion}
\label{sec:discussion}

Our analysis of the distribution of S0 galaxies in the 2D PPS of galaxy clusters regions with maximally relaxed cores provides statistically significant evidence that passive and active, SF lenticulars constitute two truly distinct subpopulations, each following different evolutionary pathways in these environments. We have shown that in the PPS diagram the differences are concentrated entirely in the radial dimension. The marginal distribution of projected radii demonstrates that passive lenticulars preferentially reside in the virialized cluster cores, whereas active S0s exhibit a sharp decline in the clusters' centermost regions and instead favor the outskirts. In contrast, the marginal distributions of LOS velocities for passive and active lenticulars are statistically indistinguishable. Estimated infall times further reinforce the radial segregation of these subpopulations, with quiescent S0s having mean infall times approximately 1 Gyr longer than their SF counterparts. However, our calculations also indicate that the average infall times for both subsets remain well below those typically expected for ancient infallers. This is largely because passive S0s dominate the lenticular population across all cluster environments, reaching fractional abundances of up to $\sim 75\%$ in the outermost region of our dataset, corresponding to recent arrivals. These findings suggest that a significant fraction of the quiescent S0s now residing in virialized cluster cores likely originated outside them.

By comparing the radial trends in systemic velocity and star formation activity of S0s with those of other galaxy types, we find that the SSFR profiles of the S0--AC class exhibit an overall amplitude comparable to that of Sa+Sb galaxies, though with notable differences in their radial behavior. A similar pattern emerges in their corresponding VDLOS profiles, where the primary differences are concentrated in the inner, virialized cluster region. Within $1\,R_{\mathrm{vir}}$, the inferred kinematics for Sa+Sb galaxies is consistent with them being relatively recent infallers, whereas for the AC lenticulars it supports the notion  that they are in a relatively advanced stage of dynamical relaxation within the cluster environment. Consequently, the shallower SSFR decline of AC lenticulars compared to their Sa+Sb counterparts with decreasing clustercentric distance suggests that they are generally more resistant to environmental quenching, at least on timescales shorter than $\sim 2$ Gyr. This finding reinforces the idea that the transformation of an infalling galaxy into a fully quiescent system is a prolonged process, requiring significant time to unfold \citep{Paccagnella2016, Finn2023, ONeil2024, Muriel2025}. For SF S0s, the prolonged persistence of star formation activity in the harsh cluster environment can be attributed to its preferential concentration in the innermost disks regions, as reported by \citet{Tous2024}.

Another notable aspect of the $\Delta$MS profiles of the PCA-based S0 classes (top row of Fig.~\ref{fig:ssfr_profs}) is that, regardless of the activity level, all begin to show signs of quenching around 1.5--$2\,R_{\mathrm{vir}}$. One possible interpretation of this reduction in the SSFR among objects that lie outside the virialized cluster cores is that they follow highly radial and eccentric orbits \citep{Gill2005}, which cause them to pass through the cluster pericenter before temporarily returning to the infall region, where they mix with more recent arrivals. Theoretically, this backsplash scenario is supported by simulations of hierarchical structure formation \citep[e.g.,][]{Ghigna1998, RamirezdeSouza1998, Mamon2004}, as well as by models explaining clustercentric gradients in star-formation rates within cold dark matter cosmogonies \citep{Balogh2000}. However, given the observed characteristics of the distributions of the various S0 subpopulations in the cluster PPS, as well as the shapes of their respective radial VDLOS profiles, the most plausible explanation for the onset of the gradual SSFR decline at such large clustercentric distances is the progressive suppression of activity in the environment of infalling groups prior to their arrival in the virialized core. In this regard, \citet{Lopes2024} have recently provided strong evidence that the group environment preprocesses its member galaxies, significantly reducing their star-formation rates before they enter the central cluster region and well in advance individually infalling galaxies see their star formation diminished. As shown by these authors, the environmental influence of infalling groups is driven by the substantial increase in the local galaxy density that these galaxy systems experience when they approach within  $\sim 2\,R_{\mathrm{vir}}$ of clusters centers.

The centerward decline of the star formation activity in the S0--ACs is, nevertheless, subtler ($\lesssim 0.2$ dex in $\Delta$MS) than the one observed in the Sa+Sb subset, and not replicated by the $\Delta$MS-based MS and GV S0 classes, which maintain roughly constant radial $\Delta$MS profiles across the whole cluster region. A similar independence of the star-formation level on clustercentric radius is observed in the most active galaxy classes used as benchmark, SF and Sc+Sd. For these latter galaxy populations, this behavior can be attributed to their sparse presence in cluster cores, combined with projection effects. However, we suspect that the apparent environmental insensitivity of active S0s arises from their inferred SSFRs, which are based on the SFRs derived by \citet{Salim2018} and reflect the average star formation activity of galaxies over the last 100 Myr. If, as suggested by the sharp decline in their relative fractions with increasing infall time (Figs.~\ref{fig:S0s_regfracs} and \ref{fig:DMS_regfracs}), active S0 transition to a quiescent state in clusters, this transformation must occur on a timescale too short to be resolved by the SFR measurements. Overall, our results support a delayed-then-rapid quenching scenario as proposed for these environments \citep[e.g.,][]{Wetzel2013,Rhee2020,Finn2023}. In this framework, the SSFRs of S0s that retain some star formation remain largely unaffected for an extended period after infall before undergoing a swift and abrupt quenching, subsequently transitioning into a less active category. Such a process would leave little to no imprint on the average SSFRs of the original MS and GV classes. Consistent with this idea, the radial trends of the $\Delta$MS profiles for both the entire S0 population and the subset of least active S0--QS galaxies, which are unaffected by this issue, do show a clear decline with decreasing clustercentric distance in their SSFRs (see the bottom-right panels of Figs.~\ref{fig:ssfr_profs} and \ref{fig:ssfr_profs_DMS}). 

The less active S0--PS and QS classes, on the other hand, show reduced VDLOS profiles that closely resemble those of the E subset in terms of amplitude and shape throughout the entire cluster region. However, although the star formation activity profiles also share a similar amplitude with those of the E galaxies, their general radial behavior diverge significantly. The reason is that the radial $\Delta$MS and SSFR profiles of the Es show a clearly distinct pattern from those of the other galaxy populations analyzed. Specifically, in the case of the Es these profiles begin to decline from the very periphery of the cluster region up to $\sim 2\,R_{\mathrm{vir}}$, and then increase monotonically between 2 and $1\,R_{\mathrm{vir}}$ to the point of eventually equaling or even slightly exceeding the SSFR levels of the passive S0 classes at the same clustercentric distance. Once inside the virial region, E galaxies resume the SSFR decline, following a trend similar to that observed in the S0--PS and QS classes. Assuming that SSFR measurements in this extremely low regime ---which essentially become upper limits \citep{Salim2016}--- do not significantly affect the inferred global radial trends, we interpret the rise in star formation activity of Es at intermediate clustercentric distances as further evidence of the preprocessing experienced by S0s. Here, galaxy mergers within infalling substructures likely play a key role, ultimately driving the morphological transformation of part of the lenticular galaxies into ellipsoidal objects.

\section{Summary and conclusions}
\label{sec:conclusions}

A subset of 14 nearby cluster regions with maximally relaxed cores was selected to examine the spatial and kinematic distributions of different S0 galaxy subpopulations within the PPS of these dense galaxy aggregations. The aim was to gain deeper insights into the environmental impact on the evolutionary pathways of this Hubble type. The clusters were chosen based on their strong, extended X-ray emission, and the presence of $R\leq R_{\mathrm{vir}}$ cores that pass some of the most stringent virialization and substructure tests in the literature. To enhance the signal-to-noise ratio and encompass a broad range of environmental conditions, the selected cluster regions were normalized using virial parameters and co-added within a projected cluster radius of $3\,R_{\mathrm{vir}}$. Approximately $25\%$ of the galaxies in this stack of cluster samples are S0s, exhibiting a wide range of activity levels. The main findings of our analysis are summarized as follows:

\begin{itemize}
\setlength\itemsep{0.5em}
\item Comparison of our PCA-based spectral classification of lenticular galaxies ---that does not distinguish between activity related to SF or central BH accretion---, with one based on classes defined from the distance to the ridge of the MS of SF galaxies, shows a close alignment between the two taxonomies, confirming previous works that demonstrate that the emission-line activity of the majority of present-day S0s comes from star formation.

\item Quiescent and SF lenticulars constitute two truly distinct subpopulations, each undergoing different evolutionary pathways in the cluster environment. In the PPS diagram their differences are entirely confined to the radial dimension, with passive systems ---the dominant S0 subpopulation across the entire cluster environment--- being preferentially located in the innermost regions, whereas active lenticulars are more commonly found in the outskirts. In contrast, the marginal distributions of LOS velocities for both subpopulations of S0s remain statistically indistinguishable.

\item The distinct distributions within the cluster PPS of the quiescent and SF classes of S0s translate into mean infall times that are approximately 1 Gyr longer for the former systems, but that, overall, are significantly shorter than those typically associated with the earliest infallers. This and the observed similarity in the radial VDLOS profiles of S0s and late-type disks beyond the virial radius suggest that a substantial fraction of lenticular galaxies currently residing in cluster cores have reached these regions via secondary infall.

\item While late-type disks and SF galaxies exhibit a pronounced inward increase in velocity dispersion within the projected virial radius of relaxed clusters ---indicative of radially elongated orbits---, S0 galaxies, regardless of activity level, display radial VDLOS profiles that tend to plateau, suggesting a more isotropic orbital distribution. This behavior supports the notion that lenticular galaxies, as a whole, represent an older cluster population having likely completed more than one orbit and undergone significant dynamical mixing within the cluster core. Even the most active S0s appear to be in a relatively advanced stage of dynamical relaxation.

\item S0 galaxies, both collectively and within certain classes, exhibit radial SSFR profiles that begin declining well beyond the central virial region. Their low orbital eccentricities, inferred from VDLOS profiles, coupled with estimated first infall times of $\gtrsim 2$ Gyr, suggest that this decline is more likely driven by preprocessing in infalling groups rather than by backsplash effects.

\item The flat or gently centerward declining radial SSFR profiles observed in the S0 galaxy classes that retain some star formation activity support the delayed-then-rapid quenching scenario proposed for cluster environments as the most plausible explanation of these trends. Additionally, our results suggest that quenching in SF S0s occurs on a very short timescale ($\lesssim 0.1$ Gyr), possibly driven by an abrupt gas stripping event on their first passage through the orbital pericenter.
\end{itemize}

Finally, we wish to emphasize that the cluster selection in this study was designed to ensure dynamical homogeneity, applying strict criteria to include only galaxy associations with virialized cores. However, our final MR cluster sample spans a relatively broad mass range ---from $\sim 0.5$ to $2.5\times 10^{15}\,M_\odot$---, which suggests that galaxies may experience somewhat different physical conditions depending on their host cluster. Moreover, the SDSS-DR12 sampling may not be perfectly uniform across all selected cluster regions. Nevertheless, the fact that the inferred VDLOS profiles precisely mirror the ordering of the SSFR profiles, where greater radial anisotropy in galaxy orbits consistently corresponds to higher star formation activity, and that this ordering aligns perfectly with the expected average ages of the stellar populations across the various galaxy subsets, strongly suggests that our results are robust and unlikely to be affected by biases that could significantly impact the above conclusions.

\begin{acknowledgements}
We acknowledge financial support from the Spanish state agency MCIN/AEI/10.13039/501100011033 and by 'ERDF A way of making Europe' funds through research grant PID2022-140871NB-C22. MCIN/AEI/10.13039/501100011033 has also provided additional support through the Center of Excellence Mar\'\i a de Maeztu's award for the Institut de Ci\`encies del Cosmos at the Universitat de Barcelona under contract CEX2019-000918-M. J.L.T.\ acknowledges support by the PRE2020-091838 grant from MCIN/AEI/10.13039/501100011033 and 'ESF Investing in your future', and by the Science and Technology Facilities Council (STFC) of the UK Research and Innovation via grant reference ST/Y002644/1. This research has used data from the DR12 of the SDSS-III survey (\url{https://www.sdss3.org/dr12/}). We appreciate the efforts of everyone involved in collecting, reducing, and processing these data, and the institutions that made the survey and its public release possible.
\end{acknowledgements}

\bibliographystyle{aa}
\bibliography{references}

\begin{onecolumn}
    
\appendix

\FloatBarrier

\section{Projected phase space and sky maps of the MR cluster regions} 
\label{app:plotsclusters}

This appendix presents plots of the 2D PPS diagrams and spatial distribution of the galaxies in each of the 14 cluster regions identified as containing the most dynamically relaxed cores (MR sample). For each cluster, the left panel displays the PPS diagram, where galaxy positions are plotted as projected clustercentric distances normalized by the virial radius of their host clusters, against absolute LOS velocities normalized by the cluster velocity dispersion. The right panels show maps of the distribution of member galaxies in right ascension and declination, extending up to three virial radii from the cluster center. In all plots, gray crosses represent non-lenticular galaxy members, red dots indicate S0--PS class members, blue dots correspond to S0--AC class members, and green dots denote S0--TR objects. In the PPS diagrams, the black curve marks the caustic of the respective cluster region. In the sky maps, black concentric circles indicate distances of one, two, and three virial radii from the cluster centers, which are identified by black crosses. Additionally, orange crosses mark the locations of the brightest cluster galaxies (BCGs) ---two if the cluster harbors a double BCG and none if no clear dominant galaxy is present---, while purple crosses indicate the peak of X-ray emission. 
\begin{figure}[!htb]
    \centering
    \includegraphics[width=\columnwidth]{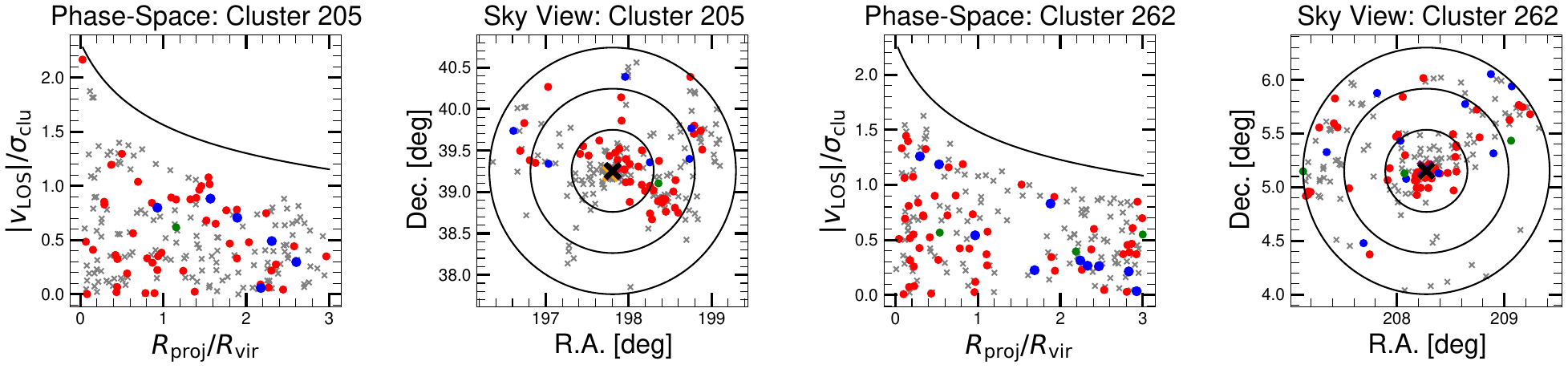}
    \includegraphics[width=\columnwidth]{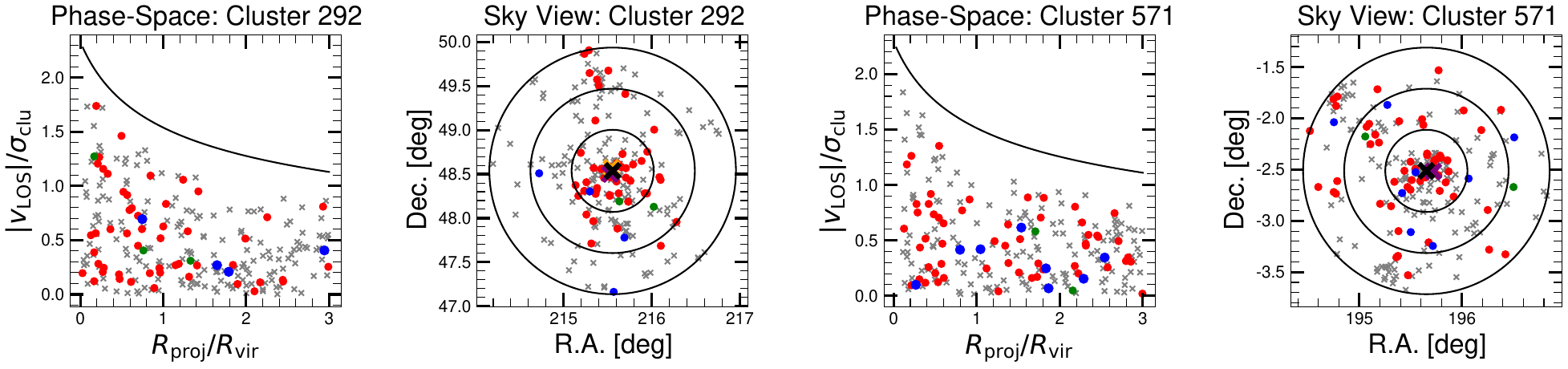}
    \includegraphics[width=\columnwidth]{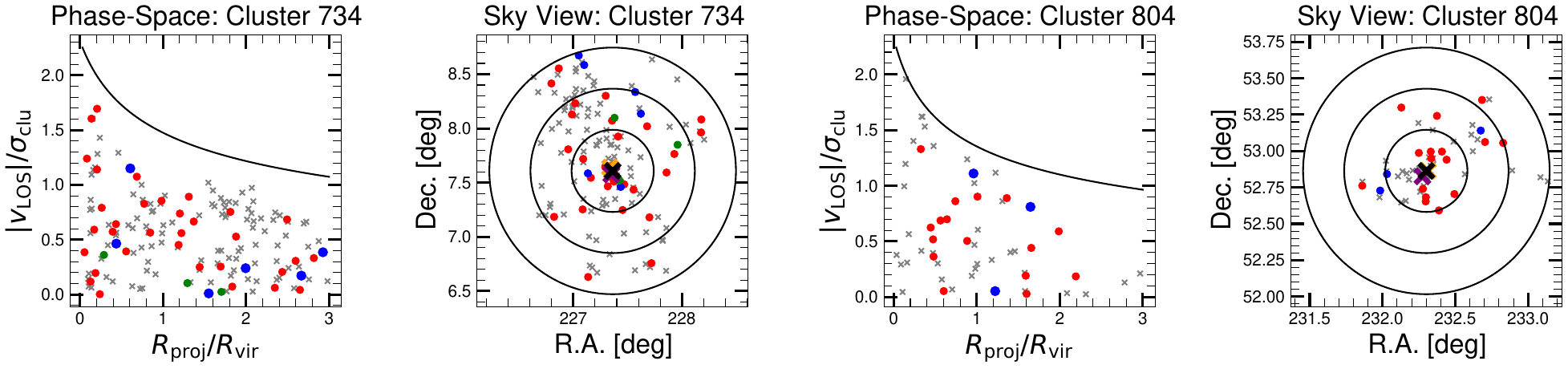}
    \caption{Projected phase space (left panels) and sky maps of the galaxy distribution (right panels) of the 14 most relaxed cluster regions analyzed in this work (MR sample). The caustic lines in the PPS diagrams have been calculated assuming that the total mass density of the clusters follows a NFW profile with a concentration parameter $C=6$. The meaning of the different symbols and curves is explained in the accompanying text.} 
    \label{fig:cont1}
\end{figure}

\begin{figure}[!htb]
    \ContinuedFloat
    \captionsetup{justification=centering}
    \centering
    \includegraphics[width=\columnwidth]{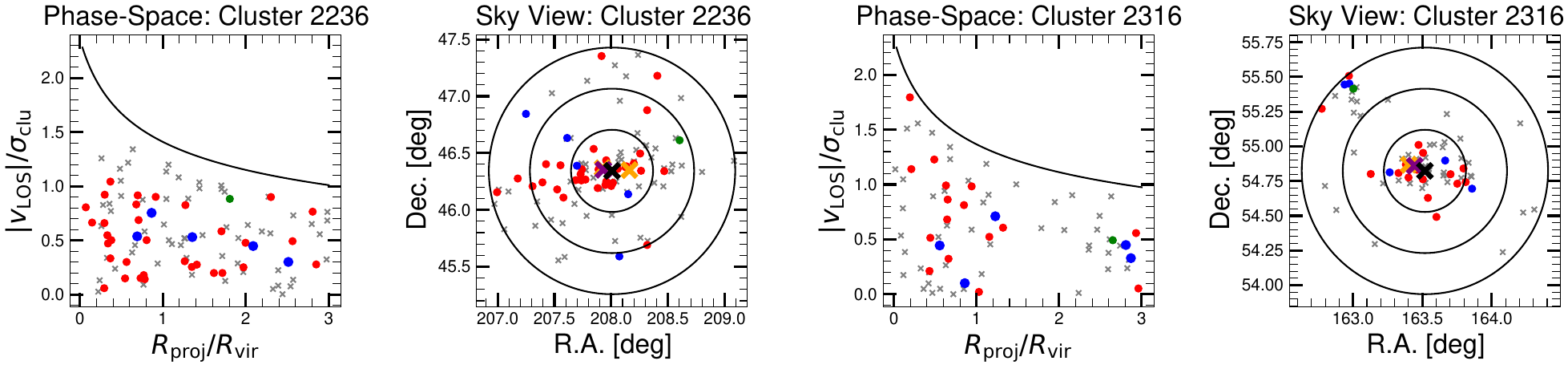}
    \includegraphics[width=\columnwidth]{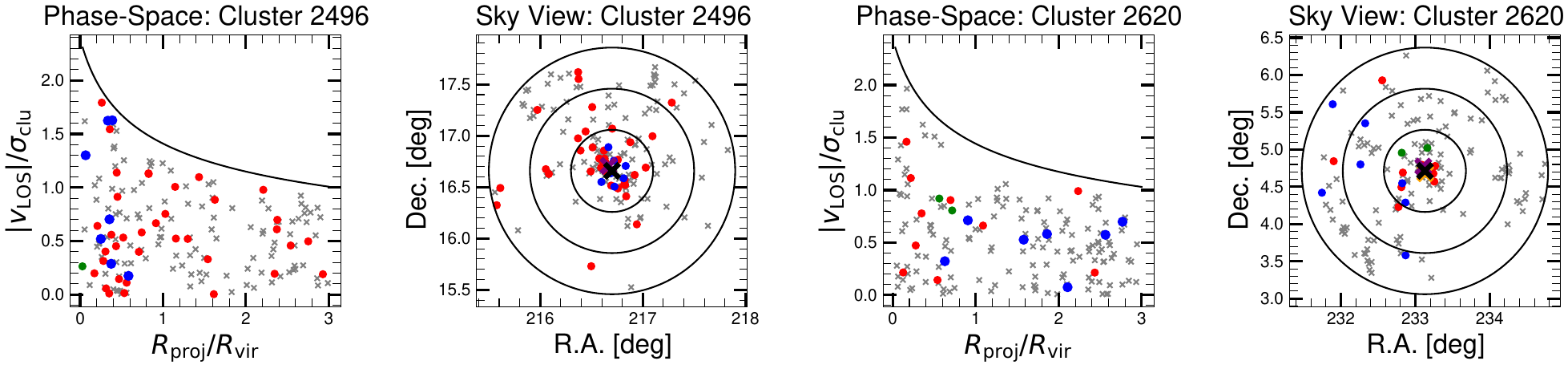}
    \includegraphics[width=\columnwidth]{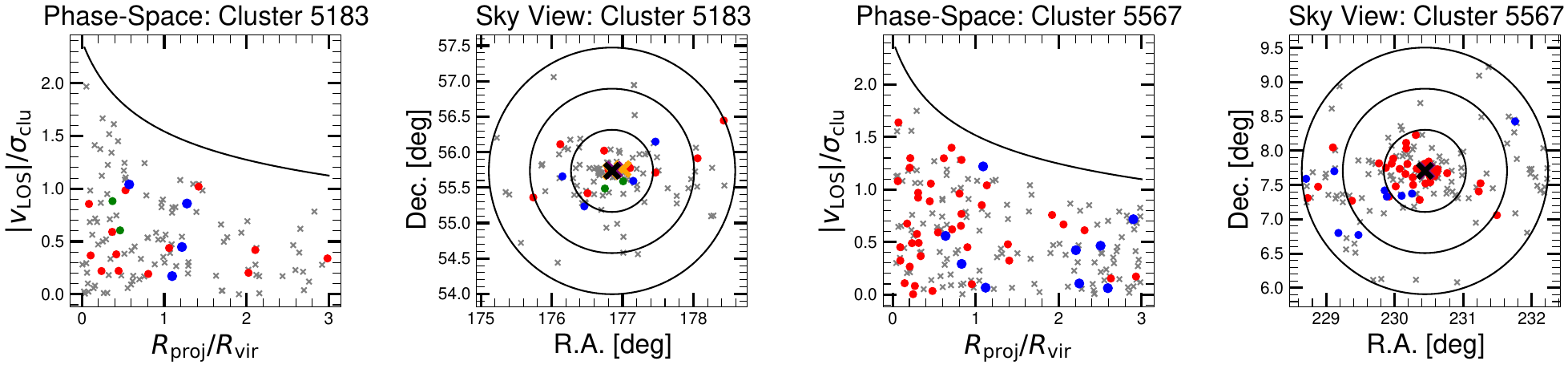}
    \includegraphics[width=\columnwidth]{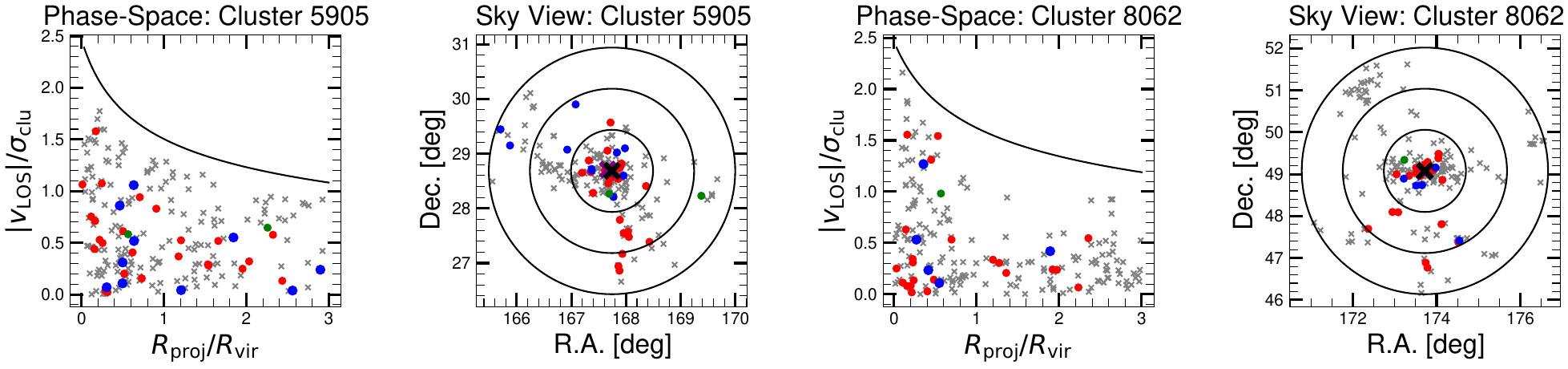}
    \caption{\textit{- Continued.}}
    \label{fig:cont2}
\end{figure}

\clearpage
\newpage
\FloatBarrier
\section{Ancillary figures}
\label{extra_material}
Additional figures that visually complement the discussions in the main text.
\begin{figure}[!htb]
    \centering
    \includegraphics[width=0.86\columnwidth]{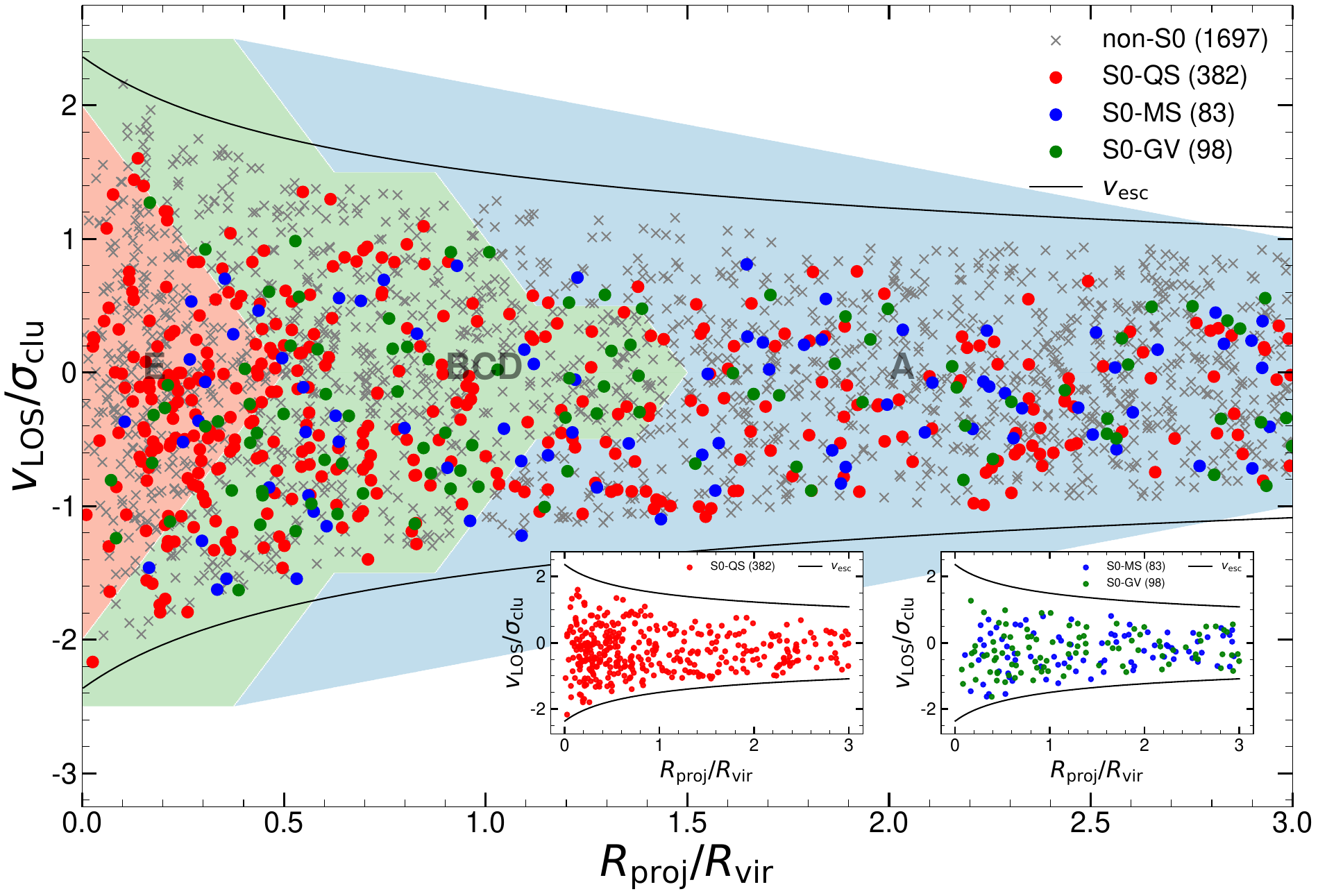}
    \caption{Same as Fig.~\ref{fig:phasespace_allgals}, but with S0 galaxies divided by their distance from the main sequence: S0--QS (red), S0--MS (blue), and S0--GV (green). The inset panels display the PPS distributions separately for the QS (left) and the combined MS+GV classes (right) from the $\Delta\mbox{MS}$ classification scheme.}
    \label{fig:phasespace_allgals_DMS}
\end{figure}

\begin{figure}[!htb]
    \centering
    \includegraphics[width=\textwidth]{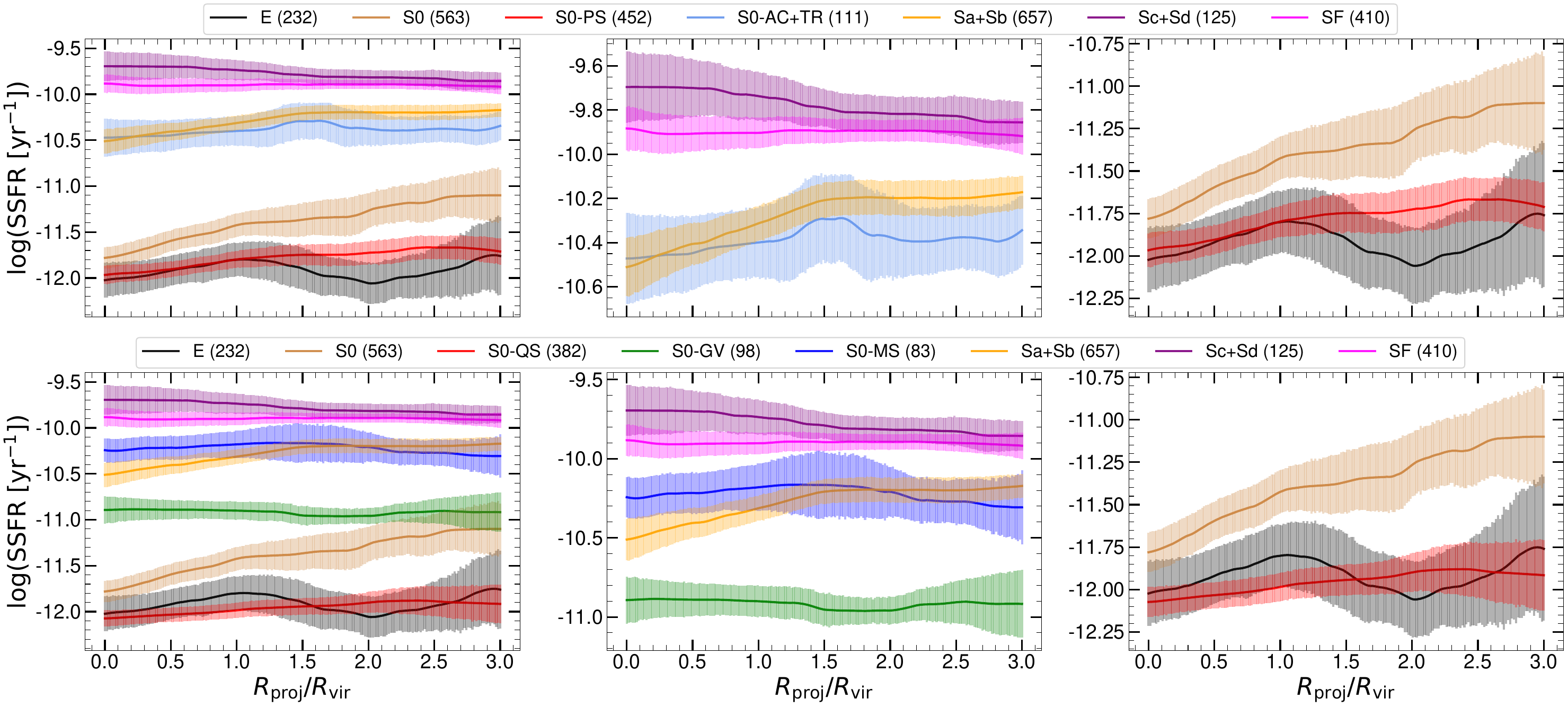}
    \caption{Same as in Fig.~\ref{fig:ssfr_profs}, but now the top row shows the radial SSFR profiles for the PCA-based classes of S0 galaxies, while the bottom row does the same for S0s classified according to the $\Delta$MS scheme. Color coding is consistent: S0--MS (blue), S0--GV (green), S0--QS (red), S0--AC+TR (light-blue), S0--PS (red), E (black), Sa+Sb (orange), Sc+Sd (purple), SF (magenta), and all S0 galaxies (brown).}
    \label{fig:ssfr_profs_DMS}
\end{figure}
\end{onecolumn}

\end{document}